\documentclass[aps,twocolumn,longbibliography,nofootinbib]{revtex4-2}

\usepackage[colorlinks=true,citecolor=blue,urlcolor=blue,linkcolor=blue,bookmarksopen]{hyperref}
\usepackage{amsmath,amssymb,bm}
\usepackage[dvips]{graphicx}
\usepackage{hyperref}
\usepackage{yfonts}
\usepackage{color}
\newcommand{\beq}{\begin{eqnarray}}
\newcommand{\eeq}{\end{eqnarray}}

\begin{document}

\title{Chiral plasma instability and inverse cascade from 
nonequilibrium left-handed neutrinos in core-collapse supernovae}

\author{Jin Matsumoto$^1$, Naoki Yamamoto$^1$, and Di-Lun Yang$^{2}$}
\affiliation{$^1$Department of Physics,  Keio University, Yokohama 223-8522, Japan \\
	$^2$Institute of Physics, Academia Sinica, Taipei, 11529, Taiwan}

\begin{abstract}
We show that the backreaction of left-handed neutrinos out of equilibrium on the matter 
sector induces an electric current proportional to a magnetic field even without 
a chiral imbalance for electrons in core-collapse supernovae.
We derive the transport coefficient of this effect based on the recently 
formulated chiral radiation transport theory for neutrinos. This chiral electric current
generates a strong magnetic field via the so-called chiral plasma instability, which 
could provide a new mechanism for the strong and stable magnetic field of magnetars.
We also numerically study the physical origin of the inverse cascade of the 
magnetic energy in the magnetohydrodynamics including this current. 
Our results indicate that incorporating the chiral effects of neutrinos would 
drastically modify the hydrodynamic evolutions of supernovae,
which may also be relevant to the explosion dynamics.
\end{abstract}
\maketitle

\section{Introduction and summary}
Following the theoretical suggestion by Lee and Yang \cite{Lee:1956qn},
Wu and collaborators experimentally discovered that the weak interaction violates 
the parity symmetry, one of the most fundamental symmetries in nature \cite{Wu:1957my}. 
In this experiment, a polarized $^{60}$Co in the external magnetic field at low temperature 
decays into $^{60}$Ni by emitting an electron (and an antineutrino) via the weak interaction. 
The observation that the electron is preferentially emitted into the direction opposite
to the magnetic field shows that the weak interaction violates parity symmetry. 
Later, it was also confirmed that neutrinos are only left-handed within 
the Standard model (SM) of particle physics \cite{Goldhaber:1958nb}.

On the other hand, this important feature of neutrinos is missed in the 
conventional radiation transport theory used in numerical simulations 
of core-collapse supernovae; see, e.g., 
Refs.~\cite{Kotake:2012nd, Burrows:2012ew, Foglizzo:2015dma, Janka:2016fox, Muller:2016izw, Radice:2017kmj}. 
Only recently, based on the idea of the relevance of the chirality of neutrinos 
in the hydrodynamic evolutions of supernovae \cite{Yamamoto:2015gzz},
the radiation transport theory incorporating this effect was 
constructed systematically from the underlying SM \cite{Yamamoto:2020zrs}. 
Using this chiral radiation transport theory, it was also found that a magnetic field 
induces nonequilibrium corrections on the neutrino energy-momentum tensor 
and neutrino number current \cite{Yamamoto:2021hjs}; 
see also Eqs.~(\ref{neutrino_current}) and (\ref{neutrino_momentum}) below. 

In this paper, we show that backreaction of these nonequilibrium left-handed 
neutrinos on the matter sector induces an electric current proportional to 
the magnetic field in core-collapse supernovae. We derive this transport coefficient
based on our previous works~\cite{Yamamoto:2020zrs, Yamamoto:2021hjs}.
As this chiral electric current generates a strong magnetic field via the so-called 
chiral plasma instability (CPI) \cite{Joyce:1997uy, Akamatsu:2013pjd}, this could 
provide a new mechanism for the strong and stable magnetic field of magnetars.
We also numerically study the physical reason why the subsequent magnetohydrodynamic 
(MHD) evolutions including this current exhibit the inverse cascade of the magnetic field.

Although neutrinos fully out of equilibrium would provide potentially dominant contributions, 
backreaction from neutrinos near equilibrium could at least qualitatively and perhaps 
even quantitatively lead to non-negligible effects on the evolution of the matter sector
as will be shown here. Whether our new mechanism operates at a more quantitative level 
should be tested by numerical applications of the chiral radiation transport theory 
for neutrinos \cite{Yamamoto:2020zrs} in the future. 
This direction would also be important to study the impacts of the chiral effects 
of neutrinos on the explosion dynamics.

Throughout this paper, we use the natural units $\hbar = c = k_{\rm B} =1$.
The electric charge $e$ is absorbed into the definition of electromagnetic fields.

\section{Chiral electric current induced by nonequilibrium neutrinos}
In Ref.~\cite{Yamamoto:2021hjs}, 
it was shown based on the chiral transport theory for neutrinos \cite{Yamamoto:2020zrs} 
that there are additional contributions from nonequilibrium neutrinos to 
the neutrino current $j_{\nu}^{i}$ and neutrino momentum density $T_{\nu}^{i0}$ 
proportional to the magnetic field ${\bm B}$:
\begin{align}
\label{neutrino_current}
\Delta j_{\nu}^{i} &= - \kappa ({\bm \nabla} \cdot {\bm v}) B^i, \\
\label{neutrino_momentum}
\Delta T_{\nu}^{i0} &= - \mu_{\nu} \kappa ({\bm \nabla} \cdot {\bm v}) B^i,
\end{align}
where $\mu_{\nu}$ is the neutrino chemical potential and ${\bm v}$ is the fluid velocity.
Here and below, $\Delta O$ denotes the contribution of chiral effects to a physical quantity $O$.
The coefficient $\kappa$ can be computed analytically under certain approximations 
as \cite{Yamamoto:2021hjs}
\beq
\kappa = \frac{1}{72 \pi M G_{\rm F}^2 (g_{\rm V}^2 + 3 g_{\rm A}^2)}
\frac{{\rm e}^{2\beta(\mu_{\rm n}-\mu_{\rm p})}}{n_{\rm n}-n_{{\rm p}}}\,,
\eeq
where $M$ is the mass of nucleons, $G_{\rm F}$ is the Fermi constant, 
$g_{\rm V,A}$ are the nucleon vector/axial charges,
$\mu$ is the chemical potential, $n$ is the number density
(with the subscripts ``{\rm p}'' and ``{\rm n}'' denoting protons and neutrons, respectively), 
and $\beta$ is the inverse temperature. 
Here, we note the relation $\Delta T_{\nu}^{i0} = \mu_{\nu} \Delta j_{\nu}^{i}$.
This can be naturally understood as the neutrino momentum density 
(which is equal to the neutrino energy current $\Delta T_{\nu}^{0i}$) 
being given by the neutrino number current multiplied by the neutrino energy 
around the Fermi surface.

From the momentum conservation law, the matter sector receives momentum kick 
from neutrinos as
\begin{align}
\Delta T_{\rm e}^{i0} = - \Delta T_{\nu}^{i0},
\end{align}
where $T_{\rm e}^{i0}$ is the momentum density of electrons. 
Here, we ignored the nucleon recoil in the neutrino scattering consistently, 
which we also used to derive Eqs.~(\ref{neutrino_current}) and 
(\ref{neutrino_momentum}) when taking the leading-order contribution in the 
expansion in terms of $|{\bm p}|/M$ (with $\bm p$ the momentum transfer) \cite{Yamamoto:2021hjs}. 
We also assume a relation between the current and momentum density
for electrons similar to the one for neutrinos above, 
$\Delta T_{\rm e}^{i0} = \mu_{\rm e} \Delta j_{\rm e}^i$, 
where $\mu_{\rm e}$ is the electron chemical potential and
$j_{\rm e}^i$ is the electron number current. 
The electric current of electrons is given by $J_{\rm e}^i = - j_{\rm e}^i$.

Combining these relations together, we obtain an electron current proportional
to the magnetic field:
\beq
\label{main}
\Delta J_{\rm e}^i = \xi_{B} B^i, 
\qquad \xi_B = -\kappa ({\bm \nabla} \cdot {\bm v})\frac{\mu_{\nu}}{\mu_{\rm e}}\,.
\eeq 
Note that this is different from the so-called chiral magnetic effect (CME) 
\cite{Vilenkin:1980fu, Nielsen:1983rb, Fukushima:2008xe} 
in that it occurs even without the chirality imbalance of electrons itself. 
Physically, this can be understood similarly to the Wu experiment showing 
the correlation between the direction of the emitted antineutrinos/electrons 
and the direction of the magnetic field. 
Our result (\ref{main}), which we have derived here for the first time based 
on the SM, can be seen as a nonequilibrium many-body manifestation of the 
microscopic parity violation. Note that in this mechanism, damping the chirality 
imbalance of electrons due to the finite electron mass is irrelevant 
unlike the scenario in Refs.~\cite{Ohnishi:2014uea, Grabowska:2014efa}.%
\footnote{One might also suspect that the nonzero neutrino mass $m_{\nu}$
can cause the chirality flipping of neutrinos.
However, such flipping is suppressed by an additional factor involving 
$m_{\nu}/\mu_{\nu} \lesssim 10^{-8}$ compared with the weak process 
without chirality flipping.}

Taking $n_{\rm n} - n_{\rm p} \sim 0.1\,{\rm fm}^{-3}$, $\mu_{\rm n} - \mu_{\rm p} \sim 100\,{\rm MeV}$, 
$\mu_{\nu} \sim \mu _{\rm e} \sim 100\,{\rm MeV}$, 
$T \sim 10\,{\rm MeV}$, $|{\bm v}| \sim 0.01$,
and the typical length scale of the system, $\ell \sim 10\,{\rm km}$, 
we have $\xi_{B} \sim 10\,{\rm MeV}$.%
\footnote{One should note that this value is sensitive to the choice of the 
parameters, and it should be rather regarded as the upper bound for $\xi_B$, 
similarly to the discussion in Ref.~\cite{Yamamoto:2021hjs}.}
While $\xi_B$ here is not associated with the chirality imbalance of electrons,
they have the same quantum number. From the comparison of Eq.~(\ref{main}) 
with the expression of the CME, 
${\bm J}_{\rm CME} = \mu_5 {\bm B}/(2\pi^2)$~\cite{Vilenkin:1980fu, Nielsen:1983rb, Fukushima:2008xe}, 
we may define an ``effective chiral chemical potential" 
$\mu_{5, {\rm eff}} = 2\pi^2 \xi_B \sim 100\,{\rm MeV}$.
Assuming $\mu_{\rm e} \gg T$, one can also relate the coefficient $\xi_B$ to 
the ``effective chiral charge" $n_{5, {\rm eff}}$ as
\begin{eqnarray}
\xi_{B} = \frac{1}{4} \biggl ( \frac{3}{\pi^4} \biggr )^{1/3} [(n_{\rm e} + n_{5, {\rm eff}})^{1/3}-(n_{\rm e} - n_{5, {\rm eff}})^{1/3}] \;, 
\nonumber \\
\label{eq: xi_B 2nd}
\end{eqnarray}
where $n_{\rm e}$ is the number density of electrons. 

In general, the backreaction leads to the modifications of the total energy current 
and charge current (except for other dissipative terms) as
\begin{eqnarray}
T^{0i}_{\rm mat}&=&(\epsilon+P)u^i-\mu_{\rm e}\xi_{B}B^i, \\
J^i &=&(n_{\rm p}-n_{\rm e})u^i+\xi_{B}B^i.
\end{eqnarray}
Here, $T^{0i}_{\rm mat}$ and $J^i$ denote the energy current of the matter sector 
and total electric current, respectively, with $\epsilon$ being 
the energy density, $P$ the pressure, and $u^{\mu}$ the fluid four velocity. 
It is however more convenient to work on the hydrodynamic equations in the 
Landau frame by redefining a fluid velocity $\tilde u^{\mu}$
(such that $T^{0i} \propto \tilde u^{i}$) as
\begin{eqnarray}
\tilde{u}^{\mu}=u^{\mu}-\frac{\mu_{\rm e}\xi_{B}}{\epsilon+P}B^{\mu}\,,
\end{eqnarray}
where $B^{\mu} = \epsilon^{\mu \nu \alpha \beta} u_{\nu} F_{\alpha \beta}/2$
is the magnetic field defined in the fluid rest frame, 
with $F_{\alpha \beta}$ as the field strength of the electromagnetic field.
Consequently, we have
\begin{eqnarray}
T^{0i}_{\rm mat}&=&(\epsilon+P)\tilde{u}^i\,, \\
J^i &=&(n_{\rm p}-n_{\rm e})\tilde{u}^i+\left[1-\frac{(n_{\rm e}-n_{\rm p})\mu_{\rm e}}{\epsilon+P}\right]\xi_{B}B^i\,.
\end{eqnarray}
When assuming the local charge neutrality $n_{\rm e}=n_{\rm p}$, we reproduce 
the result~(\ref{main}) also in the Landau frame.

\section{Helical magnetic field generated by the chiral plasma instability}
As argued in Ref.~\cite{Yamamoto:2015gzz}, any electric current of the form 
$\Delta {\bm J} = \xi_B {\bm B}$ induces the CPI \cite{Joyce:1997uy, Akamatsu:2013pjd}, 
although the origin of $\xi_B$ here is different from previous studies 
\cite{Akamatsu:2013pjd, Ohnishi:2014uea, Grabowska:2014efa, Dvornikov:2014uza, Sigl:2015xva, Yamamoto:2015gzz, Yamamoto:2021gts}. 
 
To make our paper self-contained, we here summarize several relations regarding the 
CPI; see, e.g., Ref.~\cite{Masada:2018swb}. Inserting the perturbation of the form 
$\delta {\bm B} \propto {\rm e}^{\sigma t + {\rm i}{\bm k} \cdot {\bm x}}$
(with $\sigma$ being the growth rate of the CPI) into Maxwell equations with the current 
$\Delta {\bm J} = \xi_B {\bm B}$, the linear analysis leads to the dispersion relation
\begin{eqnarray}
\sigma = \eta k (\xi_B - k) \;, 
\label{eq: dispersion relation}
\end{eqnarray}
where $k \equiv |{\bm k}|$ is the wave number.
This $\sigma$ is positive as long as $0< k <k_{\rm crit}$ with 
$k_{\rm crit} \equiv \xi_B$, and the corresponding critical length is 
\beq
\label{critical}
\lambda_{\rm crit} \equiv \frac{2\pi}{k_{\rm crit}} = \frac{2\pi}{\xi_B}.
\eeq
One also finds that $\sigma$ becomes maximum when $k = {\xi_B}/{2} \equiv k_{\rm CPI}$
and the corresponding time and length scales are 
\begin{eqnarray}
\label{typical}
\tau_{\rm CPI}=\frac{4}{ \eta \xi_B^2}\;, \quad
\lambda_{\rm CPI} \equiv \frac{2\pi}{k_{\rm CPI}} = \frac{4\pi}{\xi_B}\;,
\end{eqnarray}
respectively.

We now provide an estimate for the magnetic field generated by this CPI. 
Although the origin of $\xi_B$ is different, the argument here is similar to Ref.~\cite{Ohnishi:2014uea}.
In the ideal Fermi gas approximation, the additional energy density 
due to $\mu_{5, {\rm eff}}$ is
\beq
\Delta \epsilon = \frac{1}{4\pi^2} (\mu_{5, {\rm eff}}^4 + 6 \mu_{5, {\rm eff}}^2 \mu_{\rm e}^2)\;.
\eeq
Assuming this whole energy is converted to that of the magnetic field by the CPI, 
$B_{\rm CPI}^2/2$, one can estimate the maximum magnetic field as
\beq
B_{\rm CPI} \sim \mu_{5, {\rm eff}}^2 \sim 10^{18}\,{\rm Gauss}\,.
\eeq
We will also verify this estimate by numerical simulations of the MHD below 
(see Fig.~\ref{fig1}). Here, the strong magnetic field is generated from 
the energy temporarily stored in neutrinos, as can be seen from the relation
$\mu_{5,{\rm eff}} \propto \mu_{\nu}$. This new mechanism could potentially 
explain the origin of the gigantic magnetic field of magnetars.
Note that unlike the conventional mechanism for magnetars, the magnetic field 
generated by the CPI possesses a nonzero magnetic helicity that characterizes 
the linking structure of poloidal and toroidal magnetic fields \cite{Ohnishi:2014uea}. 
This ensures the stability of the resulting strong magnetic field.

In this estimate of the maximum magnetic field, we adopt several optimistic 
assumptions. To what extent this mechanism is efficient in core-collapse supernovae 
should be numerically checked by the chiral radiation transport theory 
for neutrinos in Ref.~\cite{Yamamoto:2020zrs}.

\section{Inverse cascade in chiral MHD}
In Ref.~\cite{Masada:2018swb}, numerical simulations of the MHD with the current 
$\Delta {\bm J} = \xi_B {\bm B}$ (chiral MHD) in the protoneutron star were performed. 
Consequently, the CPI and the subsequent inverse cascade of the magnetic field 
in the late nonlinear phase for $8 \times 10^{-4} \le \xi_{B, {\rm ini}} \le 2 \times 10^{-2}$ 
(in the units of $100\,{\rm MeV} = 1$) are observed; see also 
Refs.~\cite{Brandenburg:2017rcb, Schober:2017cdw} for the inverse cascade 
of the chiral MHD in the context of the early Universe. 
As we discussed above, in this paper we provide a new mechanism for $\xi_B$, 
which leads to a rather larger value $\xi_B \sim 0.1$. Although the chiral MHD with
this value has not been tested previously, one expects that this would also lead to 
the inverse cascade of the magnetic field.  
More generally, one can ask the physical origin of the inverse cascade of 
the magnetic field in the present system unlike the usual MHD. 
To address this question, we extend the work \cite{Masada:2018swb} in somewhat 
wider range of the initial value of $\xi_B$ including $\xi_B \sim 0.1$
($10^{-5} \le \xi_{B, {\rm ini}} \le 10^{-1}$) and clarify the origin of the inverse cascade.

The governing chiral MHD equations are given in Ref.~\cite{Masada:2018swb}.
For the present numerical simulations, we rewrite these equations into a 
conservative form using Maxwell equations as
\begin{gather} 
\frac{\partial \rho}{\partial t} + {\bm \nabla} \cdot (\rho \mbox{\boldmath $v$}) = 0 \;, 
\label{eq: mass conservation} \\
\frac{\partial}{\partial t}(\rho \mbox{\boldmath $v$}) + {\bm \nabla} \cdot \biggl [ \rho \mbox{\boldmath $v$} \mbox{\boldmath $v$} - \mbox{\boldmath $B$} \mbox{\boldmath $B$}
+ \biggl (P + \frac{B^2}{2} \biggr ) \textbf{I} \biggr ] = \mbox{\boldmath $S$} \;, 
\label{eq: momentum conservation} \\
\frac{\partial}{\partial t} \biggl ( \frac{1}{2} \rho v^2 + \frac{1}{\Gamma - 1}P + \frac{B^2}{2} \biggr ) 
+ {\bm \nabla} \cdot \biggl [ \biggl ( \frac{1}{2} \rho v^2 + \frac{\Gamma}{\Gamma - 1}P \biggr ) \mbox{\boldmath $v$} 
\nonumber \\ 
+ \mbox{\boldmath $E$} \times \mbox{\boldmath $B$} \biggr ] 
= \mbox{\boldmath $S$} \cdot \mbox{\boldmath $v$} - \Delta {\bm J} \cdot {\bm E} \;, 
\label{eq: energy conservation} \\
\frac{\partial \mbox{\boldmath $B$}}{\partial t} = 
{\bm \nabla} \times (\mbox{\boldmath $v$} \times \mbox{\boldmath $B$}) + \eta {\bm\nabla}^2 \mbox{\boldmath $B$} + \eta  \; {\bm \nabla} \times (\xi_B \mbox{\boldmath $B$}) \;, 
\label{eq: induction equation} \\
\frac{\partial n_{5, {\rm eff}}}{\partial t} = \frac{1}{2 \pi^2} \mbox{\boldmath $E$} \cdot \mbox{\boldmath $B$} \;. 
\label{eq: anomaly equation}
\end{gather}
Equations~(\ref{eq: mass conservation})--(\ref{eq: induction equation})
correspond to the mass conservation, momentum conservation, 
energy conservation, and induction equation, respectively. 
Here, $\rho$ is the rest-mass density, ${\bm E}$ is the electric field, 
$\eta$ is the resistivity, $\Gamma$ is the ratio of specific heats 
(which we assume to be the value of the ideal gas, $\Gamma = 5/3$), 
 $\textbf{I}$ is the unit matrix, and
\begin{eqnarray}
\mbox{\boldmath $S$} = \rho \nu {\bm\nabla}^2 \mbox{\boldmath $v$} + \frac{1}{3} \rho \nu {\bm \nabla} ({\bm \nabla} \cdot \mbox{\boldmath $v$}),
\end{eqnarray}
with $\nu$ as the viscosity. 
To describe the evolution of $n_{5,{\rm eff}}$, we also postulate Eq.~(\ref{eq: anomaly equation}) 
similar to the chiral anomaly relation that stands for the helicity conservation
(see, e.g., Ref.~\cite{Yamamoto:2015gzz}).
In Eq.~(\ref{eq: anomaly equation}), the advection, diffusion, chiral separation effect, 
and cross helicity are ignored for simplicity as in Ref.~\cite{Masada:2018swb}.

In fact, the total helicity conservation is derived from the spatial integration of 
Eq.~(\ref{eq: anomaly equation}) as
\begin{eqnarray}
\frac{{\rm d} H_{\rm tot}}{{\rm d} t} = 0 \;, \quad H_{\rm tot} \equiv N_{5, {\rm eff}} + \frac{H_{\rm mag}}{4 \pi^2} \;,
\end{eqnarray}
where
\begin{eqnarray}
N_{5, {\rm eff}} \equiv \int {\rm d}^3\mbox{\boldmath $x$} \ n_{5, {\rm eff}} \;, 
\label{eq: effective chiral charge}
\quad
H_{\rm mag} \equiv \int {\rm d}^3\mbox{\boldmath $x$} \ \mbox{\boldmath $A$} \cdot \mbox{\boldmath $B$} \;, 
\label{eq: magnetic helicity}
\end{eqnarray}
are the global effective chiral charge and the magnetic helicity, respectively.
Here, $\mbox{\boldmath $A$}$ is the vector potential defined by 
${\bm B} = {\bm \nabla} \times {\bm A}$.

One can eliminate the electric field in the governing equations above through 
the modified Ohm's law including the chiral current, 
\begin{eqnarray}
\mbox{\boldmath $E$} + \mbox{\boldmath $v$} \times \mbox{\boldmath $B$} = \eta (\mbox{\boldmath $J$} - \Delta \mbox{\boldmath $J$}) \;,
\label{eq: Ohm's law}
\end{eqnarray}
where ${\bm J} = {\bm \nabla} \times {\bm B}$ is the total electric current.

As for the approximate Riemann solver, the HLLD scheme \cite{Miyoshi05} is 
used in our code to solve chiral MHD equations 
(\ref{eq: mass conservation})--(\ref{eq: anomaly equation}) in a conserved form. 
We use a MUSCL-type interpolation method to attain second-order accuracy 
in space while the temporal accuracy obtains second order by using Runge-Kutta 
time integration. In addition, the constrained transport method is implemented in 
our code to guarantee the condition ${\bm \nabla} \cdot {\bm B} = 0$ \cite{Evans88}. 
Our numerical setups are almost same as those in Ref.~\cite{Masada:2018swb} 
except for several physical parameters. We adopt $\nu=0.01$ and $\eta=1$ in 
all our numerical runs. The initial value $\xi_{B, {\rm ini}}$ (listed in Table~\ref{table1}) 
varies from $10^{-1}$ to $10^{-5}$ among models. 

\begin{table}
\begin{center}
\caption{Summary of the simulation runs.}
\begin{tabular}{ccccc}
\hline
\hline
Name & $L$ & $\xi_{B,{\rm ini}}$ & $\tau_{\rm CPI}$ \\
\hline
model 1 & $8 \times 10^{2}$ & $10^{-1}$ & $4 \times 10^{2}$ \\
model 2 & $8 \times 10^{3}$ & $10^{-2}$ & $4 \times 10^{4}$ \\
model 3 & $8 \times 10^{4}$ & $10^{-3}$ & $4 \times 10^{6}$ \\
model 4 & $8 \times 10^{5}$ & $10^{-4}$ & $4 \times 10^{8}$ \\
model 5 & $8 \times 10^{6}$ & $10^{-5}$ & $4 \times 10^{10}$ \\
\hline
\hline
\label{table1}
\end{tabular}
\end{center}
\end{table}

In our numerical runs, we resolve $\lambda_{\rm crit}$ in Eq.~(\ref{critical}) by 
10 grid points and take the grid size $\Delta = \lambda_{\rm crit}/10$. 
The number of grid points in our simulations is fixed ($N^3=128^3$). However, 
the size of the calculation domain, $L= N \times \Delta$, is changed between models 
because of the variation of $\xi_{B, {\rm ini}}$ in models. The typical timescale of 
the CPI, $\tau_{\rm CPI}$, and $L$ in each model are also listed in Table~\ref{table1}.%
\footnote{As global simulations with the macroscopic length $\ell$ are computationally too 
expensive with the current numerical resources, here we instead perform local simulations 
with the size of the small domain, e.g., $L \simeq 1.6 \times 10^{-12}\,{\rm m}$ for model 1, 
by extrapolating the near-equilibrium regime for neutrinos.}


\begin{figure}
\begin{center}
\scalebox{0.8}{{\includegraphics{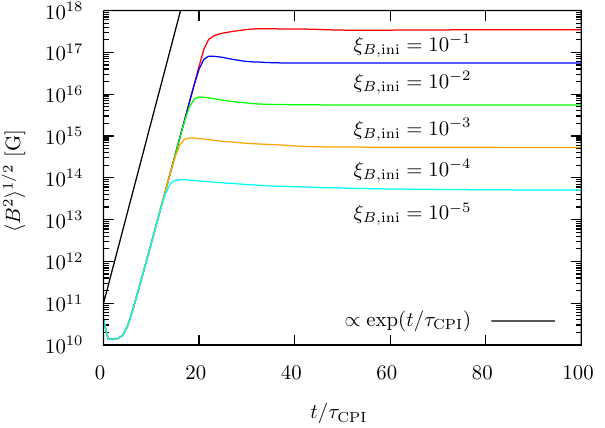}}}
\caption{Time evolution of $\langle B^2 \rangle^{1/2}$.}
\label{fig1}
\end{center}
\end{figure}

\begin{figure}
\begin{center}
\scalebox{0.8}{{\includegraphics{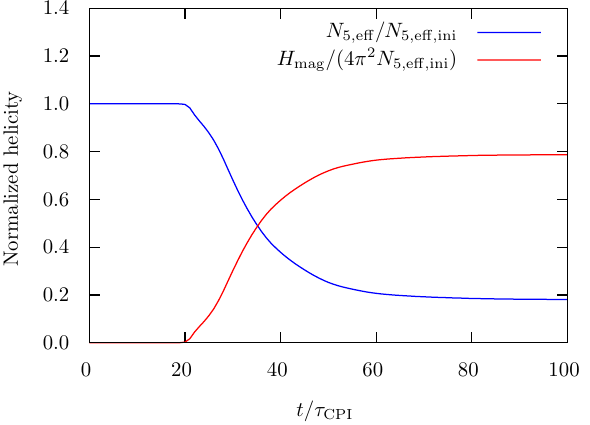}}}
\caption{Time evolution of normalized helicity in model $1$.}
\label{fig2}
\end{center}
\end{figure}

\begin{figure}
\begin{center}
\scalebox{0.112}{{\includegraphics{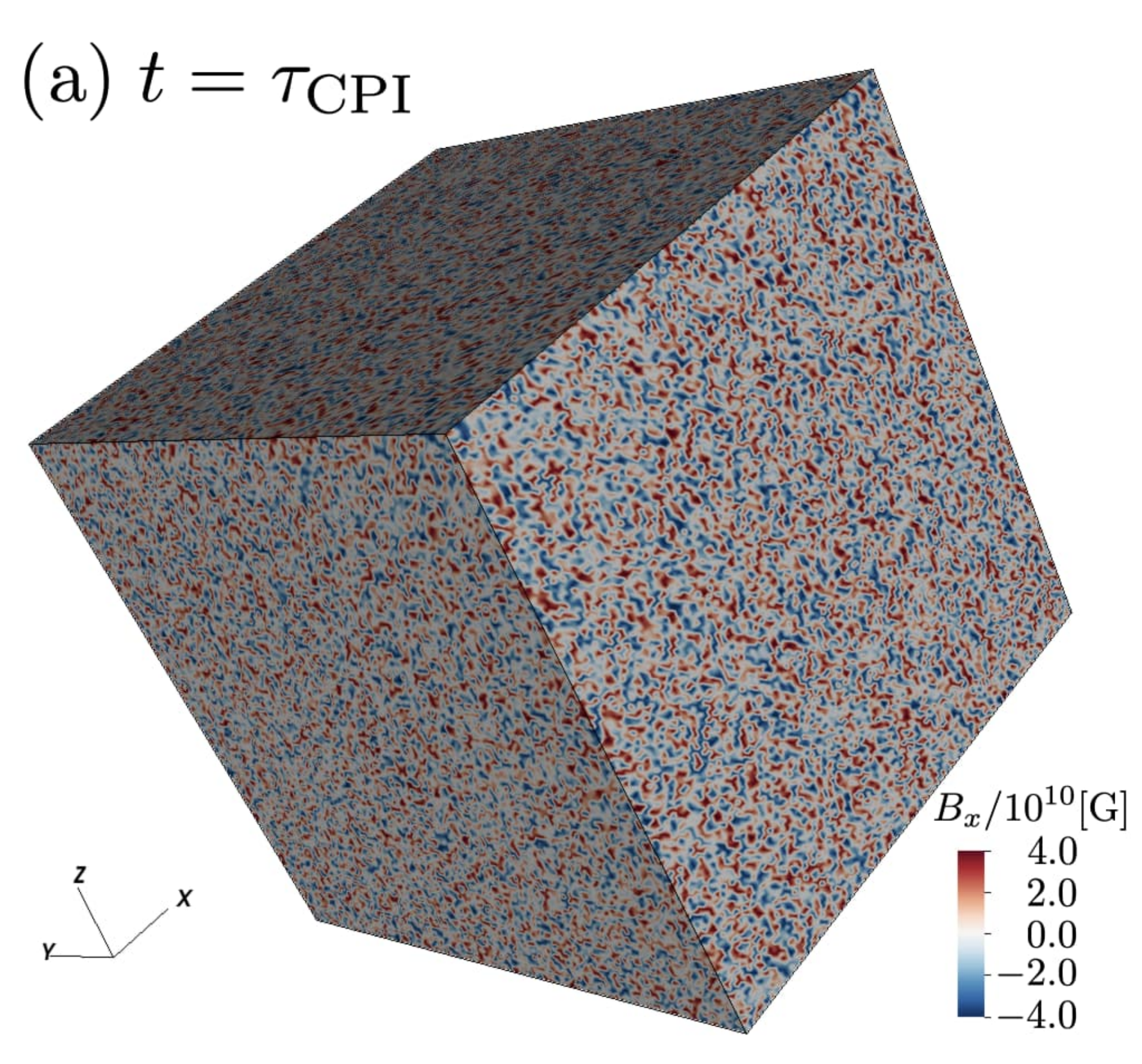}}}
\scalebox{0.112}{{\includegraphics{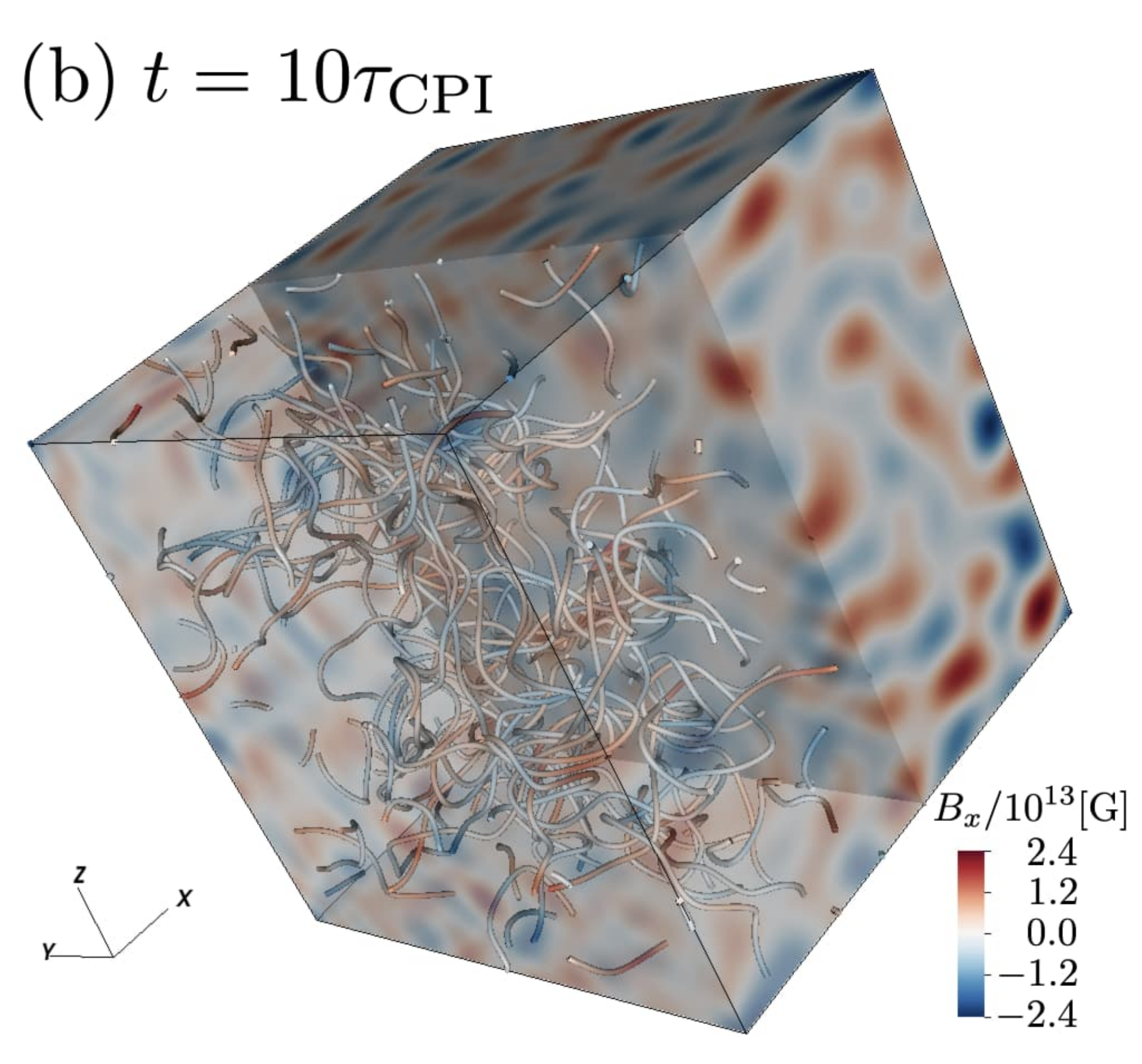}}}\\
\scalebox{0.112}{{\includegraphics{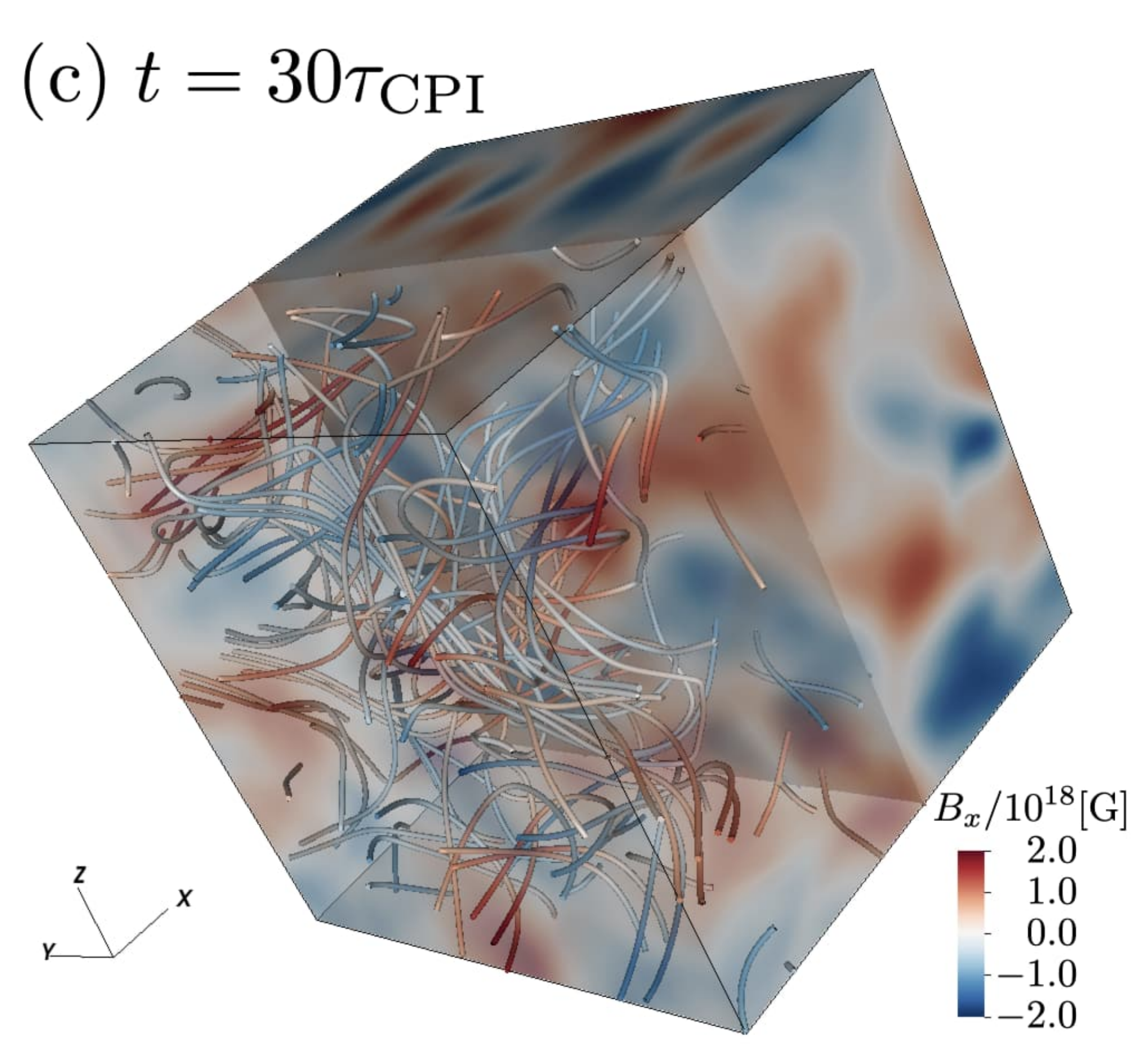}}}
\scalebox{0.112}{{\includegraphics{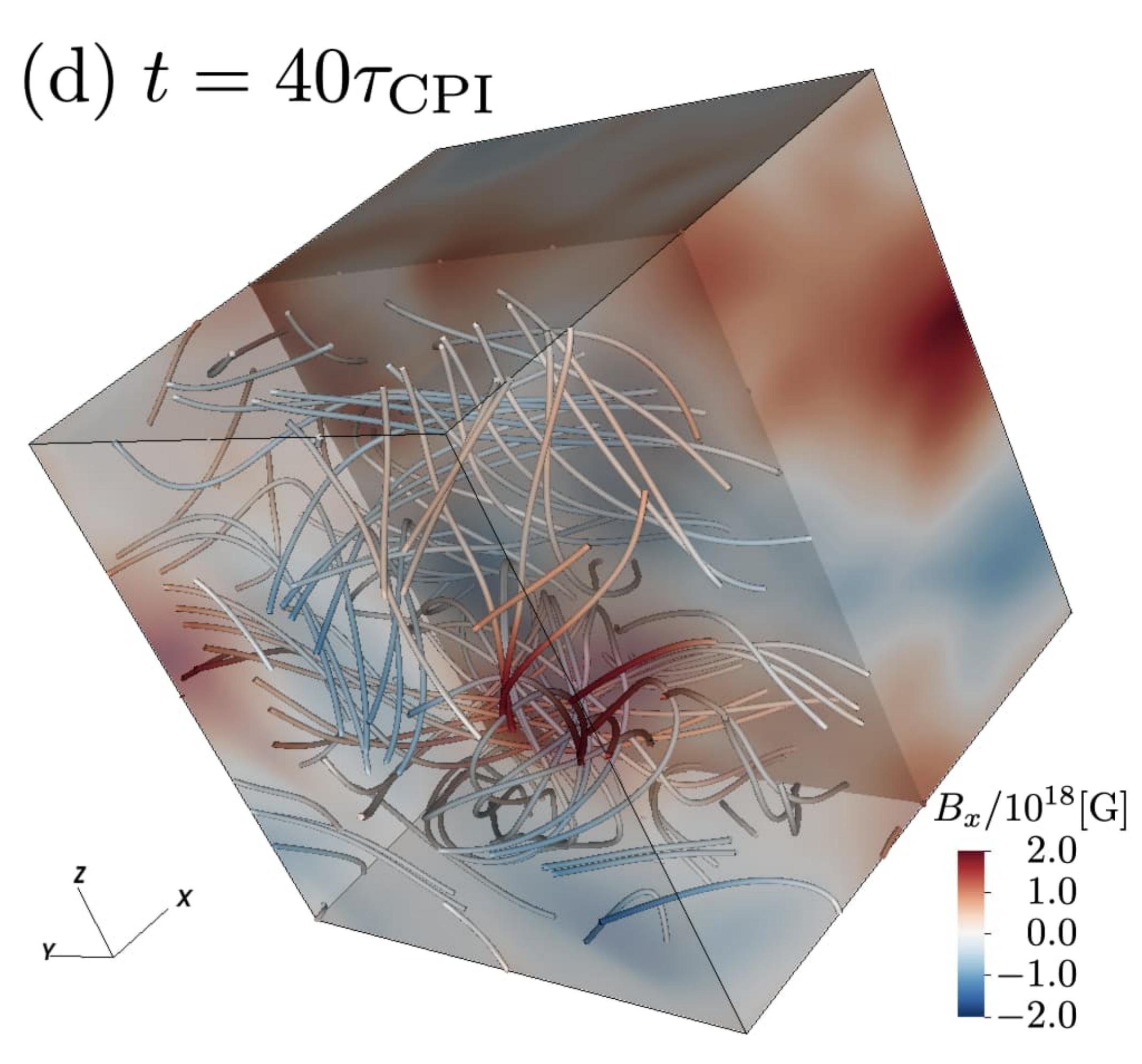}}}\\
\scalebox{0.112}{{\includegraphics{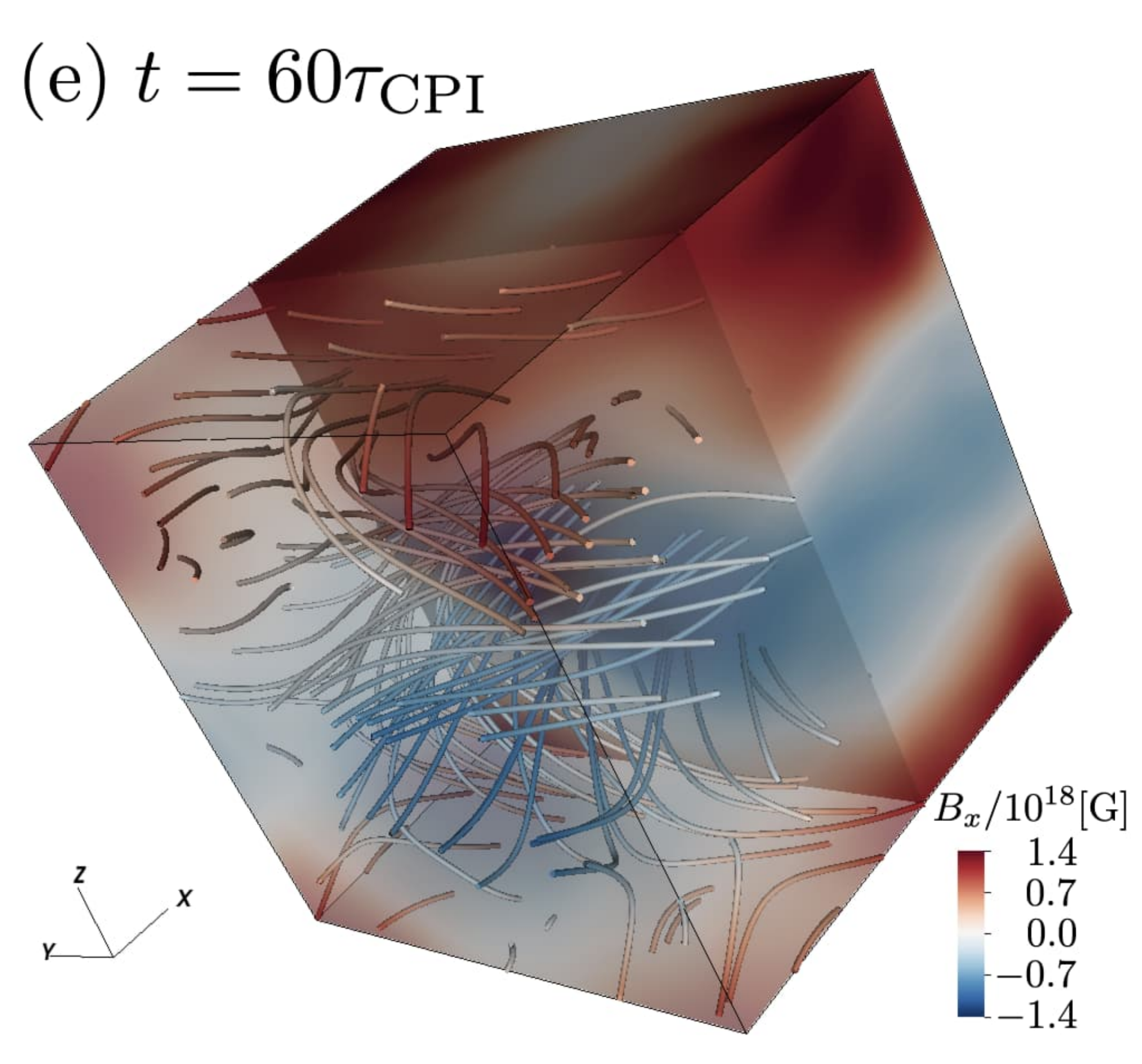}}}
\scalebox{0.112}{{\includegraphics{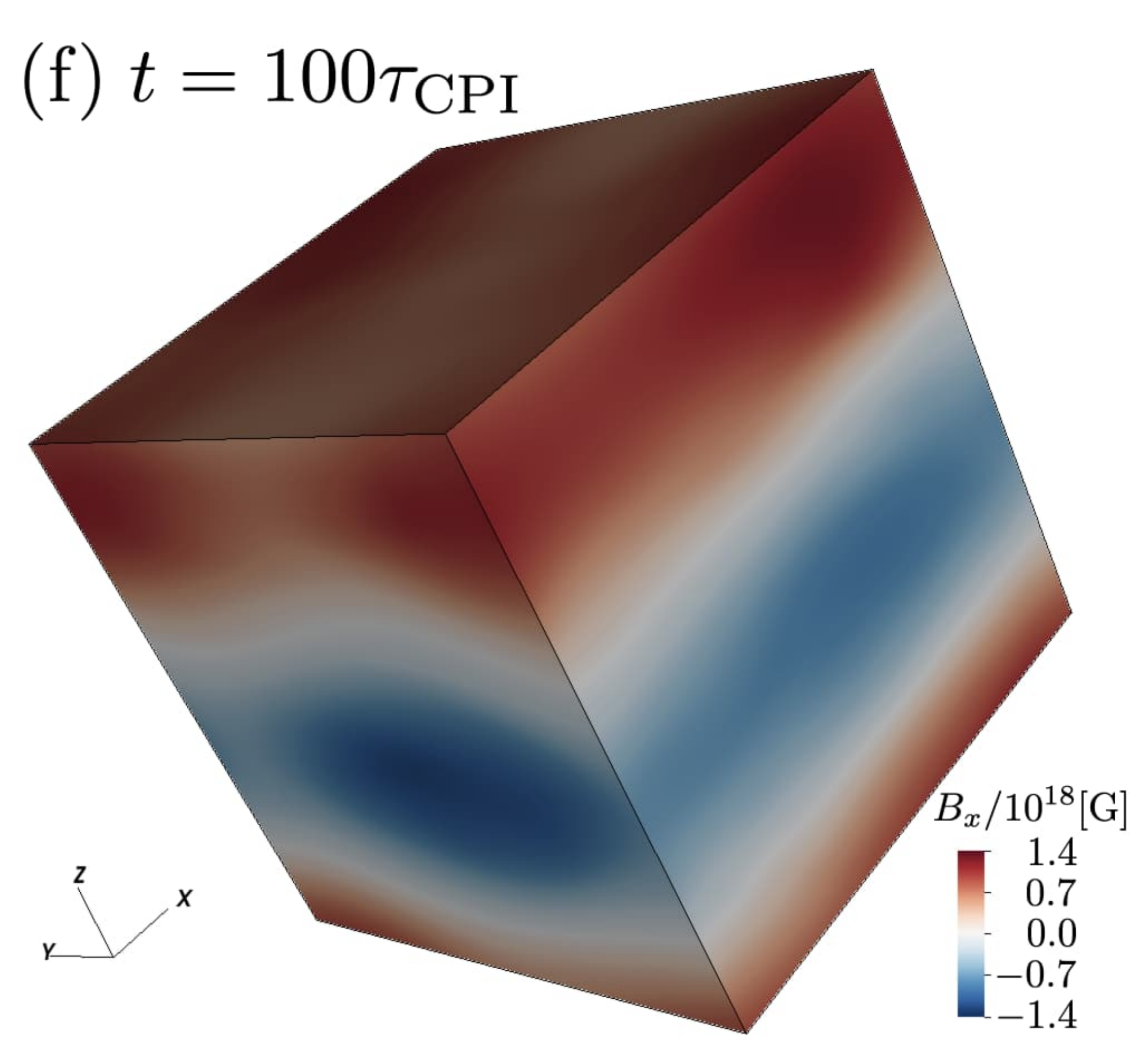}}}
\caption{Time evolution of 3D visualization of $B_x$ and magnetic field lines in model 1. 
Panels (a), (b), (c), (d), (e), and (f) correspond to the time $t/\tau_{\rm CPI}$=$1$, $10$, $30$, $40$, $60$, and $100$, respectively.}
\label{fig3}
\end{center}
\end{figure}

\begin{figure}
\begin{center}
\scalebox{0.112}{{\includegraphics{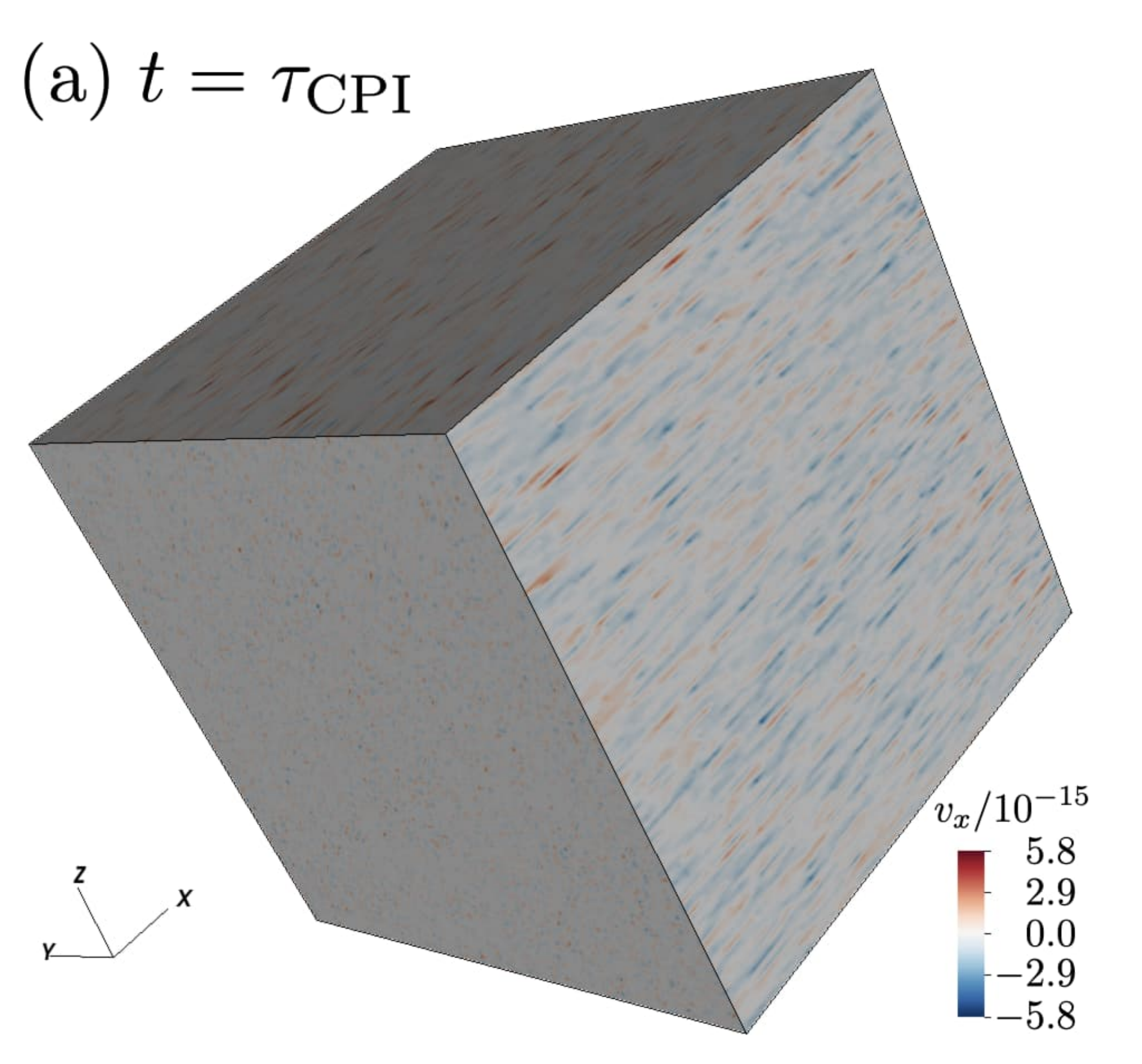}}}
\scalebox{0.112}{{\includegraphics{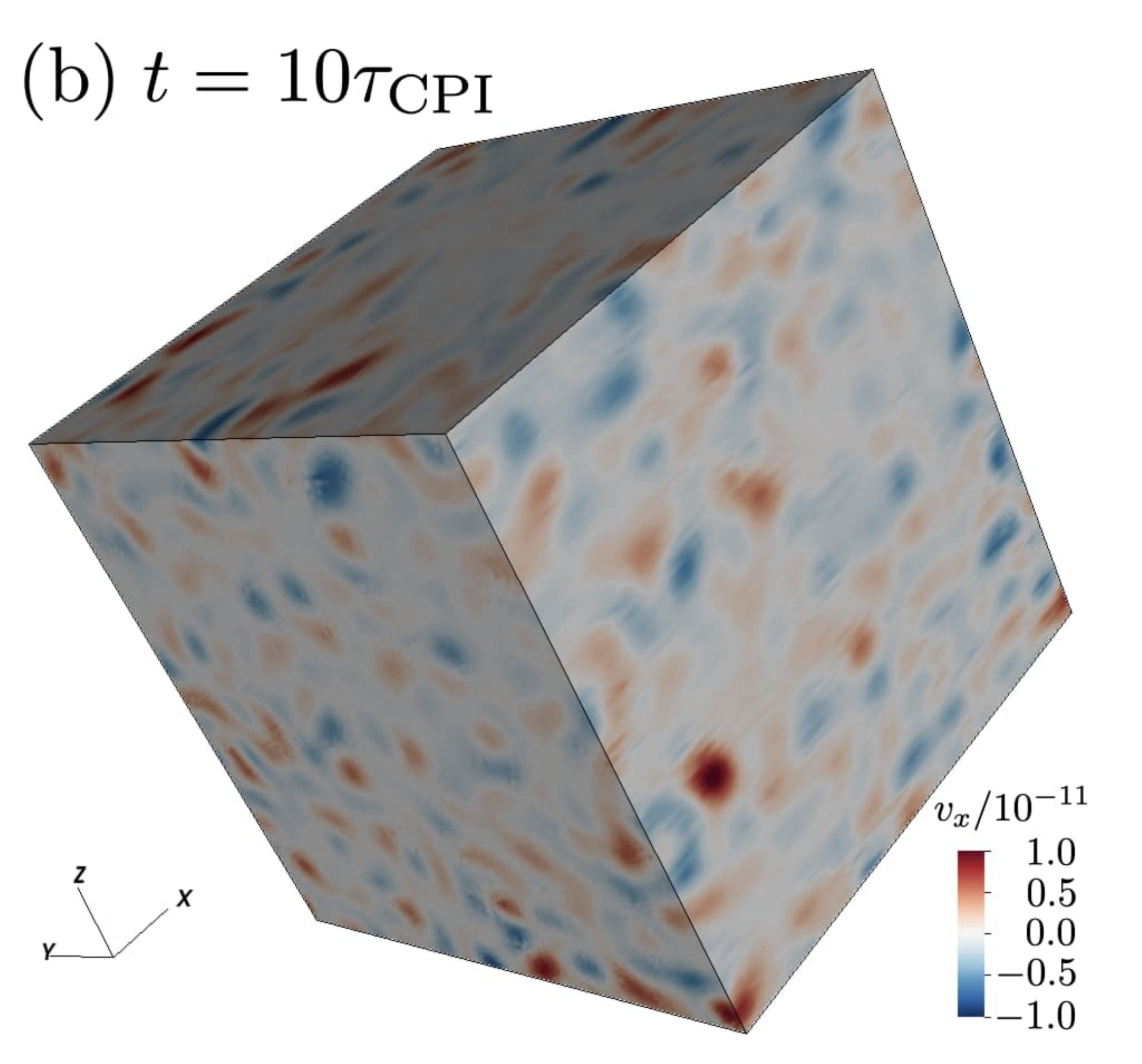}}}\\
\scalebox{0.112}{{\includegraphics{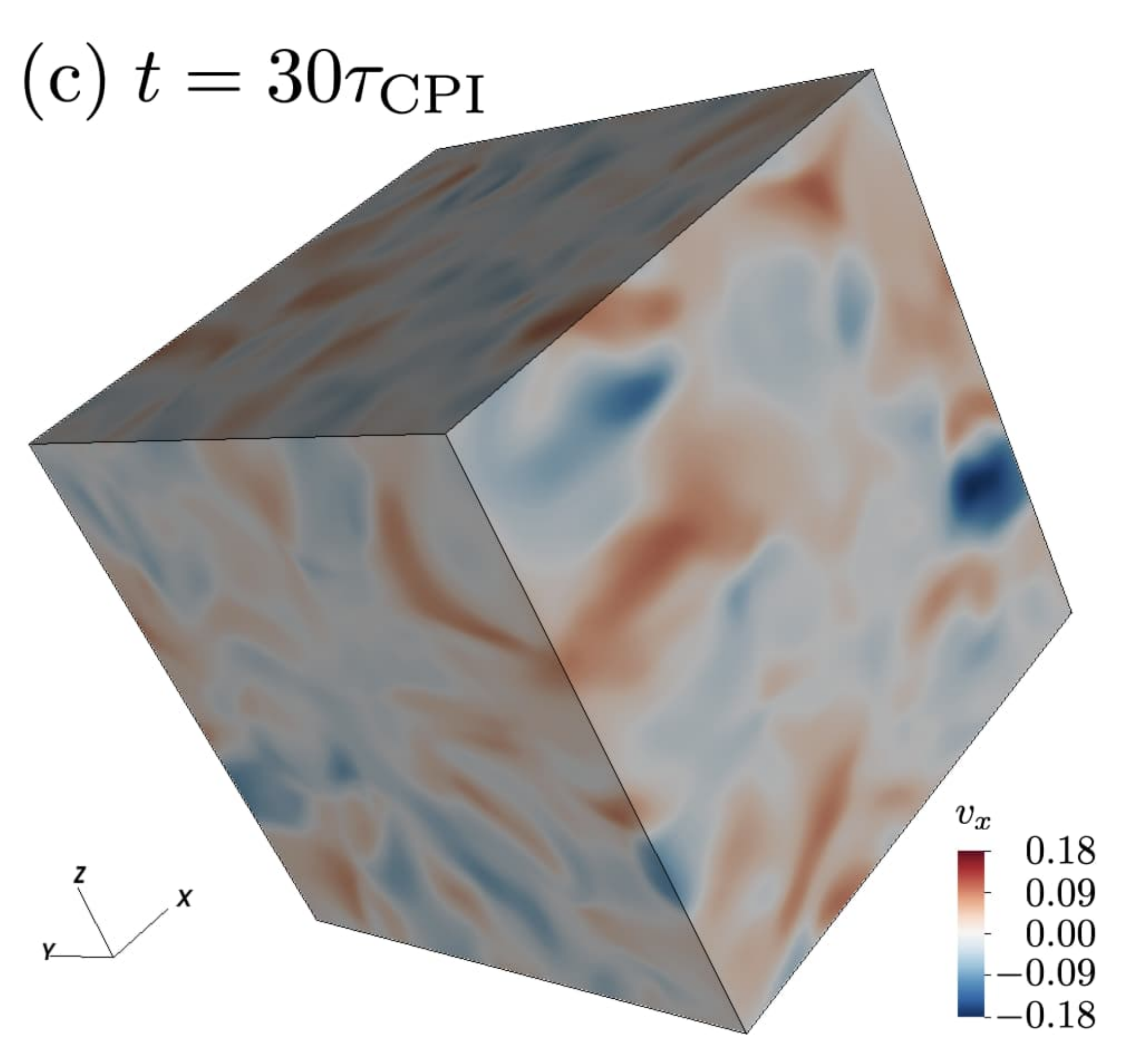}}}
\scalebox{0.112}{{\includegraphics{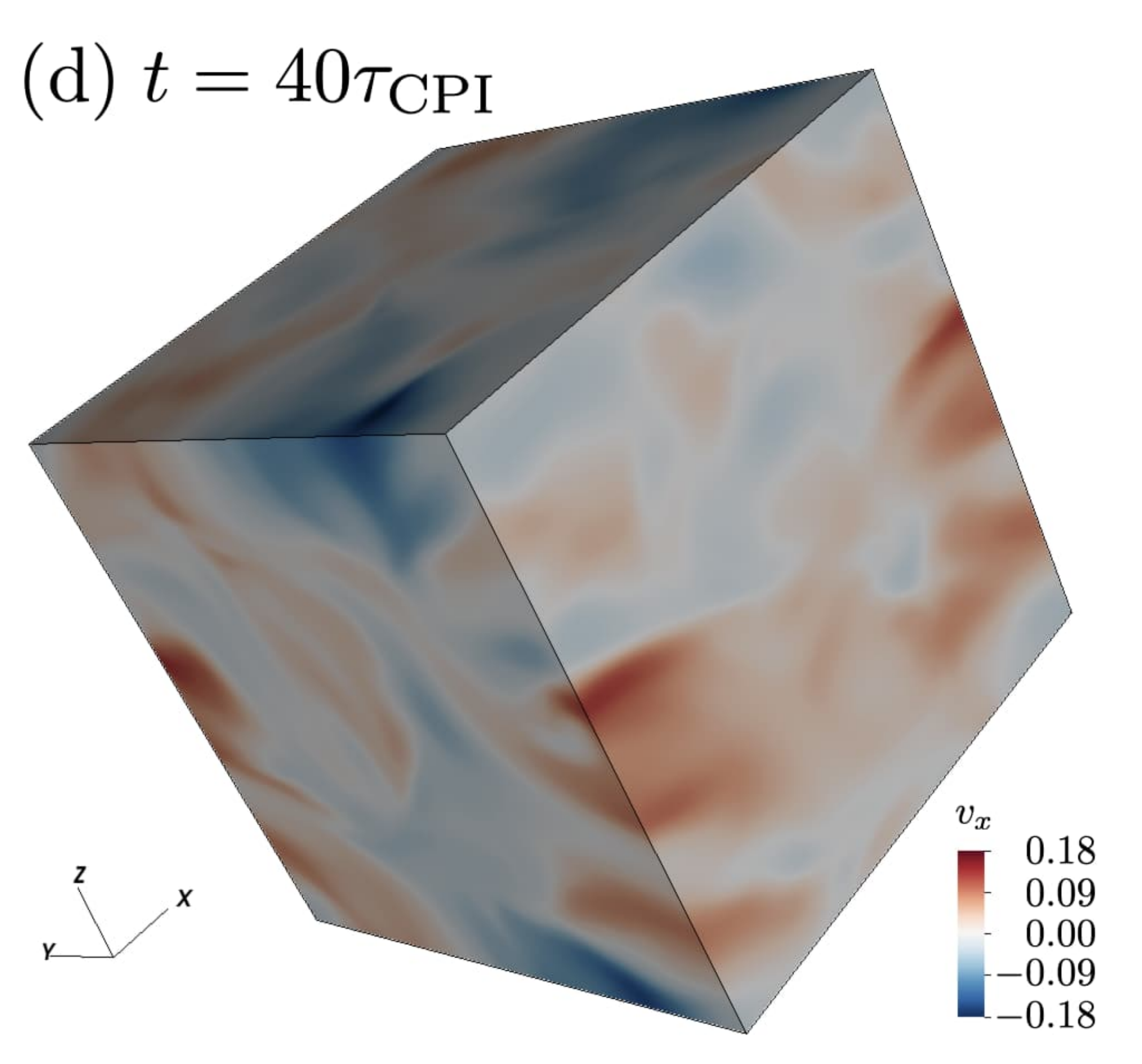}}}\\
\scalebox{0.112}{{\includegraphics{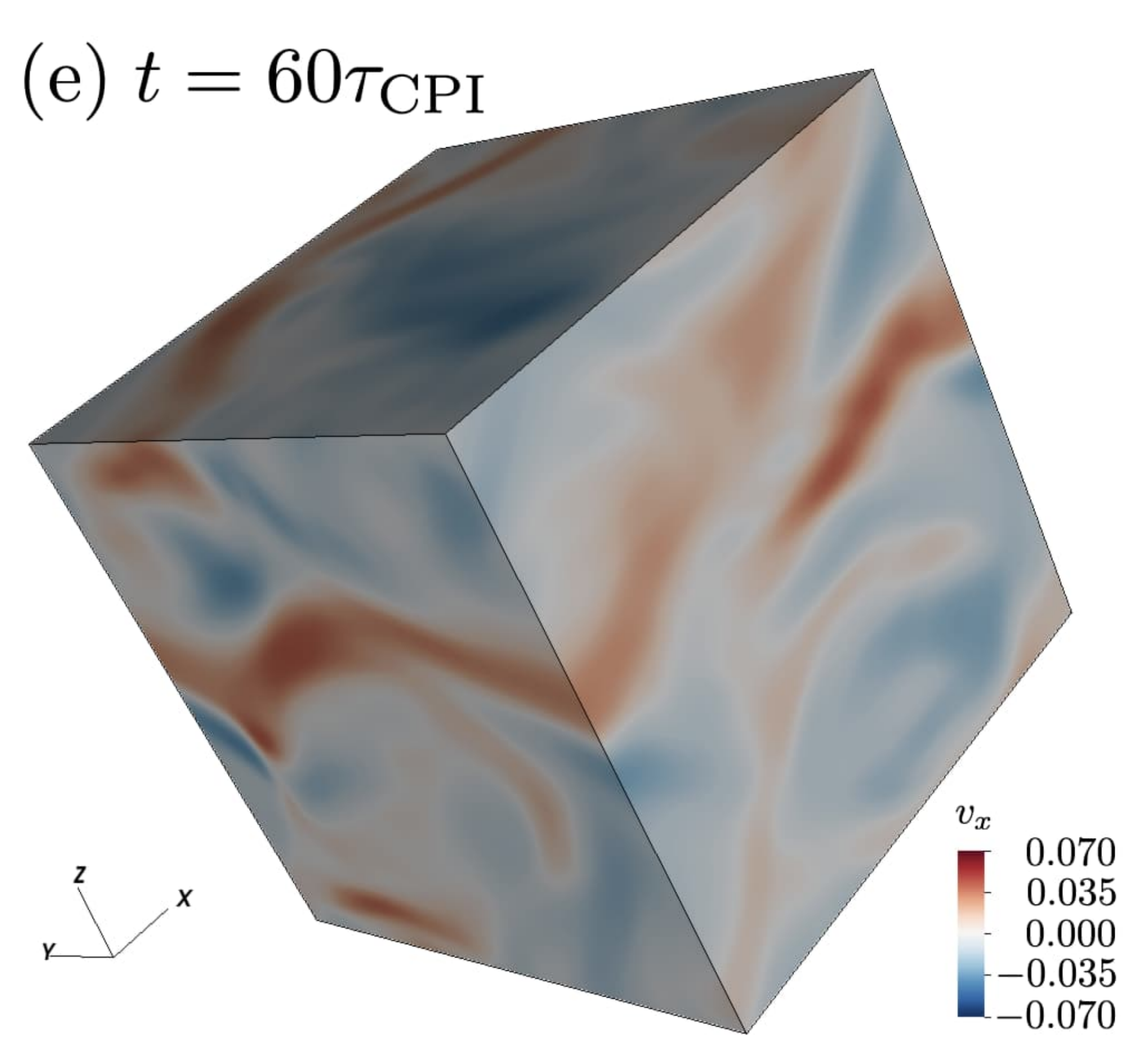}}}
\scalebox{0.112}{{\includegraphics{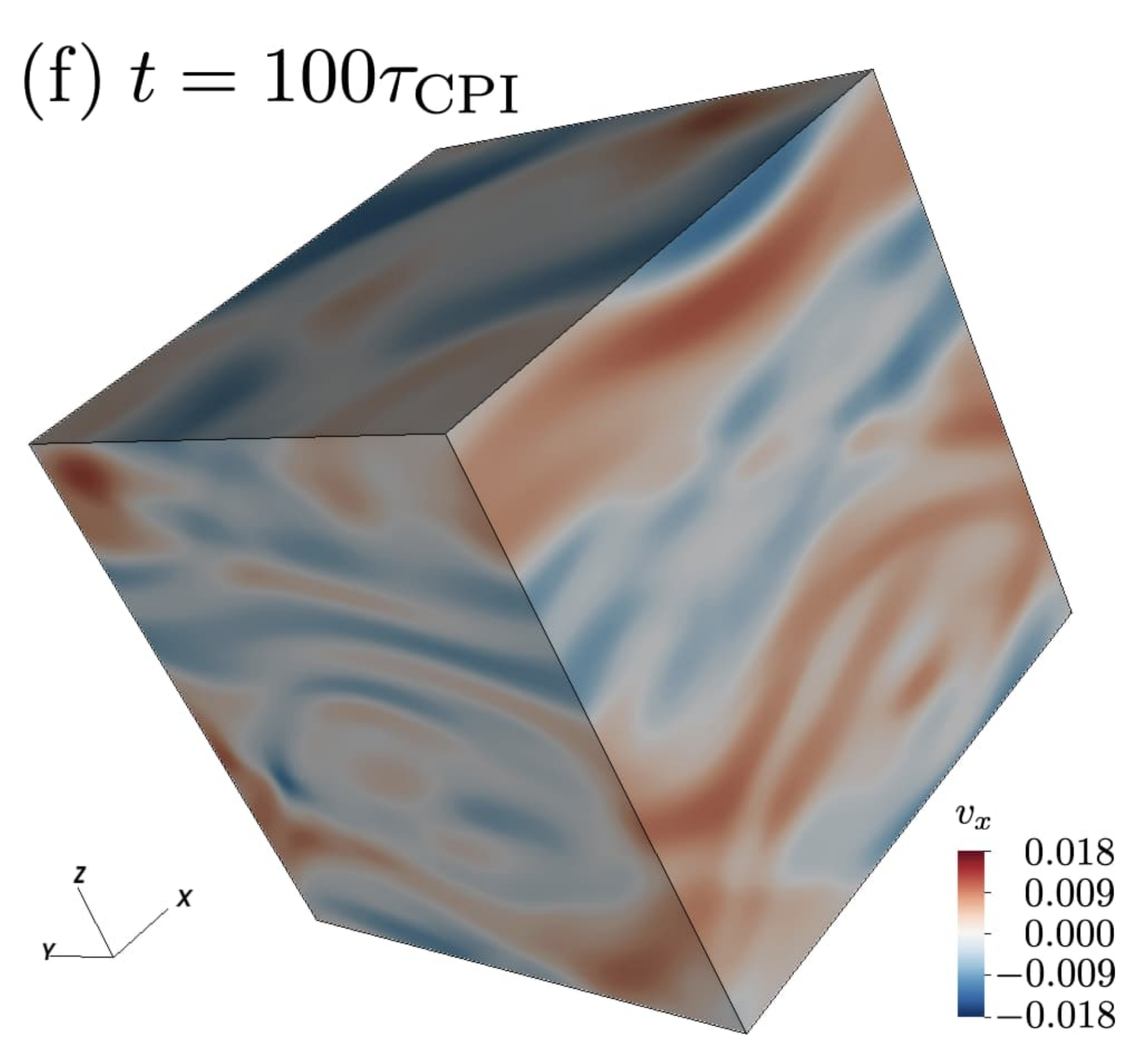}}}
\caption{Temporal evolution of 3D visualization of $v_x$ in model 1. 
Time in each panel is the same as that in Fig.~\ref{fig3}.}
\label{fig4}
\end{center}
\end{figure}

\begin{figure}
\begin{center}
\scalebox{0.8}{{\includegraphics{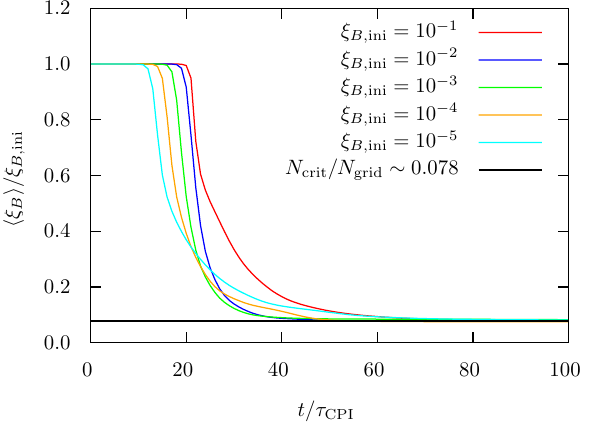}}}
\caption{Time evolution of $\langle \xi_B \rangle$. 
}
\label{fig5}
\end{center}
\end{figure}

\begin{figure*}[htbp]
\begin{center}
\scalebox{0.13}{{\includegraphics{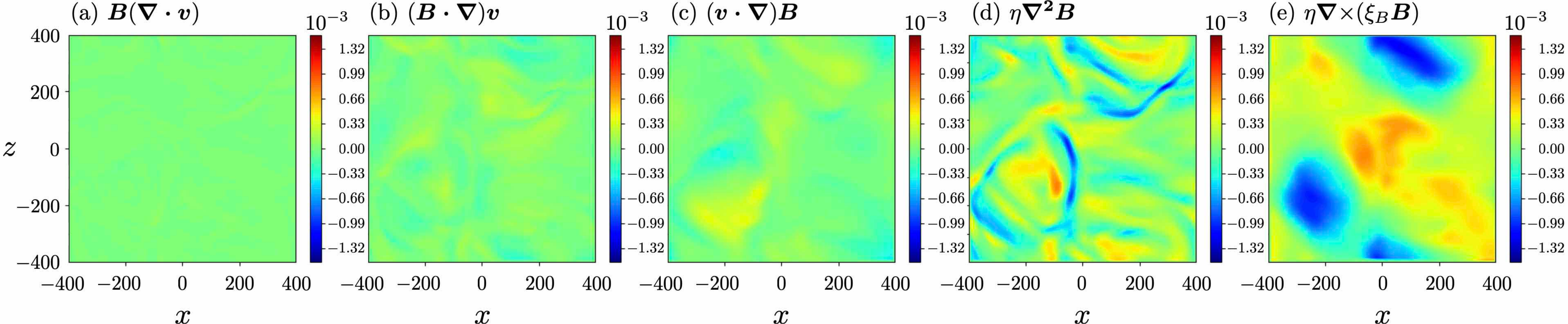}}}
\caption{Comparison of the terms in the induction equation at $t=40 \tau_{\rm CPI}$ for model 1.}
\label{fig6}
\end{center}
\end{figure*}

Figure~\ref{fig1} shows the temporal evolution of the strength of the 
volume-averaged magnetic field $\langle B^2 \rangle^{1/2}$, where 
the volume average of a physical quantity $O$ is defined by
\begin{eqnarray}
\langle O \rangle = \frac{1}{V} \int {\rm d}^3\mbox{\boldmath $x$} \ O \;, \quad
V \equiv \int {\rm d}^3\mbox{\boldmath $x$} \;.
\end{eqnarray}
The analytically predicted temporal evolution of the magnetic field due to the 
growth of the CPI ($\propto {\rm e}^{t/\tau_{\rm CPI}}$) is also shown by a 
black solid line, for reference. In all models, the growth rate of the magnetic field 
in the early phase of simulation runs after the relaxation of the initial perturbation 
of the magnetic field has good agreement with the linear analysis of the CPI.
In the later phase of the simulations, the amplification of the magnetic field 
due to the CPI is saturated. 

The magnetic helicity $H_{\rm mag}$ is drastically generated in the nonlinear phase. 
The temporal evolutions of $H_{\rm mag}$ and $N_{5, {\rm eff}}$ for 
model 1 are plotted by red and blue lines in Fig.~\ref{fig2}. 
The vertical axis is normalized by the initial effective chiral charge, $N_{5, {\rm eff,ini}}$. 
From the point of view of the conservation of the total helicity, the decrease of 
$N_{5, {\rm eff}}$ compensates for the increase of $H_{\rm mag}$. As discussed later, 
the saturation level of $N_{5, {\rm eff}}$, which is related to $\xi_B$ by 
Eq.~(\ref{eq: xi_B 2nd}), depends on $L$.

The inverse cascade of the magnetic energy and fluid kinetic energy is also observed 
in all our models. Figure~\ref{fig3} shows several time snapshots of 3D rendering 
of $B_x$ and magnetic field lines in the whole calculation domain of model 1 
($\xi_{B, {\rm ini}}=10^{-1}$) listed in Table~\ref{table1}. 
In Figs.~\ref{fig3}(a) and \ref{fig3}(f), only $B_x$ is illustrated. In Figs.~\ref{fig3}(b)--\ref{fig3}(e), 
the magnetic field lines are visualized in the negative $x$ region and 
the colors of the magnetic field lines indicate the strength of $B_x$. 
The 2D distribution of $B_x$ on the surface of the calculation 
box in the negative $x$ region is made transparent to depict the magnetic field lines. 
We can see that the correlation length of $B_x$ (the size of the red 
and blue regions) becomes larger over time. The final length scale 
of $B_x$ becomes comparable to $L$ as shown in Fig.~\ref{fig3}(f). 
This feature can be checked in all models listed in Table~\ref{table1}.
In addition, the inverse cascade of the fluid kinetic energy in model 1 is 
confirmed in Fig.~\ref{fig4}. The box size and time in each panel in this figure 
are the same as those in Fig.~\ref{fig3}. While $v_x$ has small-scale structures 
in the linear phase [Fig.~\ref{fig4}(b)], it gradually organizes large-scale structures 
in the nonlinear phase [Figs.~\ref{fig4}(c)--\ref{fig4}(f)].

Let us now discuss the physical reason why the magnetic field exhibits 
the inverse cascade in this system. In the linear phase of our simulation runs, 
the CPI with the typical wavelength $\lambda_{\rm CPI}$ is developed
as shown in Fig.~\ref{fig3}(b). In this phase, the anomaly equation (\ref{eq: anomaly equation}) 
shows that $n_{5,{\rm eff}}$ is independent of time and so is $\xi_B$ [see Eq.~(\ref{eq: xi_B 2nd})]. 
This is confirmed in Fig.~\ref{fig5}, which shows the temporal evolution of $\langle \xi_B \rangle$.

On the other hand, $\xi_B$ decreases in the nonlinear phase of the simulation runs. 
From Eq.~(\ref{typical}), this results in the increase of $\lambda_{\rm CPI}$. 
This is the origin of the inverse cascade of the magnetic field. 
Here, as previously reported in Ref.~\cite{Masada:2018swb},
we expect that the decrease of $\xi_B$ ends when $\lambda_{\rm crit} \sim L$. 
In Fig.~\ref{fig5}, $(2 \pi/ L)/\xi_{B, {\rm ini}}$ is also shown by a solid black line. 
As we expected, all color lines approach asymptotically the black line after the linear phase. 
In this way, we can explain the inverse cascade of the magnetic field 
by the process of the CPI.  

Let us look into this mechanism more closely.
The evolution of the magnetic field is governed by the induction equation (\ref{eq: induction equation}), 
where the second and third terms on the right-hand side are related to the CPI. 
Here, we evaluate the contribution of the first term to the evolution of 
the magnetic field. This nonlinear interaction between the fluid and 
magnetic field is divided into three terms, 
\begin{eqnarray}
{\bm \nabla} \times (\mbox{\boldmath $v$} \times \mbox{\boldmath $B$}) = 
\mbox{\boldmath $B$} ({\bm \nabla} \cdot \mbox{\boldmath $v$})
+(\mbox{\boldmath $B$} \cdot {\bm \nabla}) \mbox{\boldmath $v$}
-(\mbox{\boldmath $v$} \cdot {\bm \nabla}) \mbox{\boldmath $B$}\;, \nonumber \\
\end{eqnarray}
which correspond to the compression, stretching, and advection terms, respectively. 
In Fig.~\ref{fig6}, 2D distributions of the $y$ component of the compression [Fig.~\ref{fig6}(a)], 
stretching [Fig.~\ref{fig6}(b)], advection [Fig.~\ref{fig6}(c)], diffusion [Fig.~\ref{fig6}(d)] 
and CME [Fig.~\ref{fig6}(e)] in the induction equation on the $x$-$z$ plane 
at $y=0$, when $t/\tau_{\rm CPI} = 40$, for model 1 are illustrated. 
We can see that the first three terms corresponding to the nonlinear interaction 
are relatively smaller than the last two terms. This indicates that the evolution of 
the magnetic field in our system is described mostly by the diffusion and CME, 
which is the process of the CPI.

The condition that the process of the CPI is dominant in the evolution of the 
magnetic field is derived from the induction equation (\ref{eq: induction equation})
with a generic current of the form $\Delta {\bm J} = \xi_B {\bm B}$ 
[not limited to Eq.~(\ref{main})] as
\beq
\label{condition}
|{\bm v}| \ll \eta |\xi_B|\,.
\eeq
This condition is indeed satisfied for Eq.~(\ref{main}) with the parameter choice 
above and $\eta = 1$.
Note that this is only the sufficient condition for the inverse cascade.
When this condition is not satisfied, then the nonlinear interaction between 
the fluid and magnetic field is no longer negligible. It would be interesting to
study whether the inverse cascade persists in such a case.
Finally, in this paper we performed chiral MHD simulations for several given 
values of $\xi_B$, but it would also be important to perform self-consistent 
simulations directly using the form of $\xi_B$ in Eq.~(\ref{main}).

\acknowledgments
We thank Y.~Masada and I.~Shovkovy for useful and stimulating discussions.
Numerical computations were carried out on Cray XC50 at the Center for Computational Astrophysics,
National Astronomical Observatory of Japan and on Yukawa-21 at YITP in Kyoto University.
This work was supported by the Keio Institute of Pure and Applied Sciences 
(KiPAS) project at Keio University, JSPS KAKENHI Grants No.~19K03852 and No.~22H01223 and 
the Ministry of Science and Technology, Taiwan under Grant No.~MOST 110-2112-M-001-070-MY3.

\bibliography{neutrino_ref}

\begin{thebibliography}{27}%
\makeatletter
\providecommand \@ifxundefined [1]{%
 \@ifx{#1\undefined}
}%
\providecommand \@ifnum [1]{%
 \ifnum #1\expandafter \@firstoftwo
 \else \expandafter \@secondoftwo
 \fi
}%
\providecommand \@ifx [1]{%
 \ifx #1\expandafter \@firstoftwo
 \else \expandafter \@secondoftwo
 \fi
}%
\providecommand \natexlab [1]{#1}%
\providecommand \enquote  [1]{``#1''}%
\providecommand \bibnamefont  [1]{#1}%
\providecommand \bibfnamefont [1]{#1}%
\providecommand \citenamefont [1]{#1}%
\providecommand \href@noop [0]{\@secondoftwo}%
\providecommand \href [0]{\begingroup \@sanitize@url \@href}%
\providecommand \@href[1]{\@@startlink{#1}\@@href}%
\providecommand \@@href[1]{\endgroup#1\@@endlink}%
\providecommand \@sanitize@url [0]{\catcode `\\12\catcode `\$12\catcode
  `\&12\catcode `\#12\catcode `\^12\catcode `\_12\catcode `\%12\relax}%
\providecommand \@@startlink[1]{}%
\providecommand \@@endlink[0]{}%
\providecommand \url  [0]{\begingroup\@sanitize@url \@url }%
\providecommand \@url [1]{\endgroup\@href {#1}{\urlprefix }}%
\providecommand \urlprefix  [0]{URL }%
\providecommand \Eprint [0]{\href }%
\providecommand \doibase [0]{https://doi.org/}%
\providecommand \selectlanguage [0]{\@gobble}%
\providecommand \bibinfo  [0]{\@secondoftwo}%
\providecommand \bibfield  [0]{\@secondoftwo}%
\providecommand \translation [1]{[#1]}%
\providecommand \BibitemOpen [0]{}%
\providecommand \bibitemStop [0]{}%
\providecommand \bibitemNoStop [0]{.\EOS\space}%
\providecommand \EOS [0]{\spacefactor3000\relax}%
\providecommand \BibitemShut  [1]{\csname bibitem#1\endcsname}%
\let\auto@bib@innerbib\@empty
\bibitem [{\citenamefont {Lee}\ and\ \citenamefont {Yang}(1956)}]{Lee:1956qn}%
  \BibitemOpen
  \bibfield  {author} {\bibinfo {author} {\bibfnamefont {T.~D.}\ \bibnamefont
  {Lee}}\ and\ \bibinfo {author} {\bibfnamefont {C.-N.}\ \bibnamefont {Yang}},\
  }\bibfield  {title} {\bibinfo {title} {{Question of Parity Conservation in
  Weak Interactions}},\ }\href {https://doi.org/10.1103/PhysRev.104.254}
  {\bibfield  {journal} {\bibinfo  {journal} {Phys. Rev.}\ }\textbf {\bibinfo
  {volume} {104}},\ \bibinfo {pages} {254} (\bibinfo {year}
  {1956})}\BibitemShut {NoStop}%
\bibitem [{\citenamefont {Wu}\ \emph {et~al.}(1957)\citenamefont {Wu},
  \citenamefont {Ambler}, \citenamefont {Hayward}, \citenamefont {Hoppes},\
  and\ \citenamefont {Hudson}}]{Wu:1957my}%
  \BibitemOpen
  \bibfield  {author} {\bibinfo {author} {\bibfnamefont {C.~S.}\ \bibnamefont
  {Wu}}, \bibinfo {author} {\bibfnamefont {E.}~\bibnamefont {Ambler}}, \bibinfo
  {author} {\bibfnamefont {R.~W.}\ \bibnamefont {Hayward}}, \bibinfo {author}
  {\bibfnamefont {D.~D.}\ \bibnamefont {Hoppes}},\ and\ \bibinfo {author}
  {\bibfnamefont {R.~P.}\ \bibnamefont {Hudson}},\ }\bibfield  {title}
  {\bibinfo {title} {{Experimental Test of Parity Conservation in $\beta$
  Decay}},\ }\href {https://doi.org/10.1103/PhysRev.105.1413} {\bibfield
  {journal} {\bibinfo  {journal} {Phys. Rev.}\ }\textbf {\bibinfo {volume}
  {105}},\ \bibinfo {pages} {1413} (\bibinfo {year} {1957})}\BibitemShut
  {NoStop}%
\bibitem [{\citenamefont {Goldhaber}\ \emph {et~al.}(1958)\citenamefont
  {Goldhaber}, \citenamefont {Grodzins},\ and\ \citenamefont
  {Sunyar}}]{Goldhaber:1958nb}%
  \BibitemOpen
  \bibfield  {author} {\bibinfo {author} {\bibfnamefont {M.}~\bibnamefont
  {Goldhaber}}, \bibinfo {author} {\bibfnamefont {L.}~\bibnamefont
  {Grodzins}},\ and\ \bibinfo {author} {\bibfnamefont {A.~W.}\ \bibnamefont
  {Sunyar}},\ }\bibfield  {title} {\bibinfo {title} {{Helicity of Neutrinos}},\
  }\href {https://doi.org/10.1103/PhysRev.109.1015} {\bibfield  {journal}
  {\bibinfo  {journal} {Phys. Rev.}\ }\textbf {\bibinfo {volume} {109}},\
  \bibinfo {pages} {1015} (\bibinfo {year} {1958})}\BibitemShut {NoStop}%
\bibitem [{\citenamefont {Kotake}\ \emph {et~al.}(2012)\citenamefont {Kotake},
  \citenamefont {Sumiyoshi}, \citenamefont {Yamada}, \citenamefont {Takiwaki},
  \citenamefont {Kuroda}, \citenamefont {Suwa},\ and\ \citenamefont
  {Nagakura}}]{Kotake:2012nd}%
  \BibitemOpen
  \bibfield  {author} {\bibinfo {author} {\bibfnamefont {K.}~\bibnamefont
  {Kotake}}, \bibinfo {author} {\bibfnamefont {K.}~\bibnamefont {Sumiyoshi}},
  \bibinfo {author} {\bibfnamefont {S.}~\bibnamefont {Yamada}}, \bibinfo
  {author} {\bibfnamefont {T.}~\bibnamefont {Takiwaki}}, \bibinfo {author}
  {\bibfnamefont {T.}~\bibnamefont {Kuroda}}, \bibinfo {author} {\bibfnamefont
  {Y.}~\bibnamefont {Suwa}},\ and\ \bibinfo {author} {\bibfnamefont
  {H.}~\bibnamefont {Nagakura}},\ }\bibfield  {title} {\bibinfo {title}
  {{Core-Collapse Supernovae as Supercomputing Science: a status report toward
  6D simulations with exact Boltzmann neutrino transport in full general
  relativity}},\ }\href {https://doi.org/10.1093/ptep/pts009} {\bibfield
  {journal} {\bibinfo  {journal} {PTEP}\ }\textbf {\bibinfo {volume} {2012}},\
  \bibinfo {pages} {01A301} (\bibinfo {year} {2012})},\ \Eprint
  {https://arxiv.org/abs/1205.6284} {arXiv:1205.6284 [astro-ph.HE]}
  \BibitemShut {NoStop}%
\bibitem [{\citenamefont {Burrows}(2013)}]{Burrows:2012ew}%
  \BibitemOpen
  \bibfield  {author} {\bibinfo {author} {\bibfnamefont {A.}~\bibnamefont
  {Burrows}},\ }\bibfield  {title} {\bibinfo {title} {{Colloquium: Perspectives
  on core-collapse supernova theory}},\ }\href
  {https://doi.org/10.1103/RevModPhys.85.245} {\bibfield  {journal} {\bibinfo
  {journal} {Rev. Mod. Phys.}\ }\textbf {\bibinfo {volume} {85}},\ \bibinfo
  {pages} {245} (\bibinfo {year} {2013})},\ \Eprint
  {https://arxiv.org/abs/1210.4921} {arXiv:1210.4921 [astro-ph.SR]}
  \BibitemShut {NoStop}%
\bibitem [{\citenamefont {Foglizzo}\ \emph {et~al.}(2015)\citenamefont
  {Foglizzo} \emph {et~al.}}]{Foglizzo:2015dma}%
  \BibitemOpen
  \bibfield  {author} {\bibinfo {author} {\bibfnamefont {T.}~\bibnamefont
  {Foglizzo}} \emph {et~al.},\ }\bibfield  {title} {\bibinfo {title} {{The
  explosion mechanism of core-collapse supernovae: progress in supernova theory
  and experiments}},\ }\href {https://doi.org/10.1017/pasa.2015.9} {\bibfield
  {journal} {\bibinfo  {journal} {Publ. Astron. Soc. Austral.}\ }\textbf
  {\bibinfo {volume} {32}},\ \bibinfo {pages} {e009} (\bibinfo {year}
  {2015})},\ \Eprint {https://arxiv.org/abs/1501.01334} {arXiv:1501.01334
  [astro-ph.HE]} \BibitemShut {NoStop}%
\bibitem [{\citenamefont {Janka}\ \emph {et~al.}(2016)\citenamefont {Janka},
  \citenamefont {Melson},\ and\ \citenamefont {Summa}}]{Janka:2016fox}%
  \BibitemOpen
  \bibfield  {author} {\bibinfo {author} {\bibfnamefont {H.~T.}\ \bibnamefont
  {Janka}}, \bibinfo {author} {\bibfnamefont {T.}~\bibnamefont {Melson}},\ and\
  \bibinfo {author} {\bibfnamefont {A.}~\bibnamefont {Summa}},\ }\bibfield
  {title} {\bibinfo {title} {{Physics of Core-Collapse Supernovae in Three
  Dimensions: a Sneak Preview}},\ }\href
  {https://doi.org/10.1146/annurev-nucl-102115-044747} {\bibfield  {journal}
  {\bibinfo  {journal} {Ann. Rev. Nucl. Part. Sci.}\ }\textbf {\bibinfo
  {volume} {66}},\ \bibinfo {pages} {341} (\bibinfo {year} {2016})},\ \Eprint
  {https://arxiv.org/abs/1602.05576} {arXiv:1602.05576 [astro-ph.SR]}
  \BibitemShut {NoStop}%
\bibitem [{\citenamefont {M\"uller}(2016)}]{Muller:2016izw}%
  \BibitemOpen
  \bibfield  {author} {\bibinfo {author} {\bibfnamefont {B.}~\bibnamefont
  {M\"uller}},\ }\bibfield  {title} {\bibinfo {title} {{The Status of
  Multi-Dimensional Core-Collapse Supernova Models}},\ }\href
  {https://doi.org/10.1017/pasa.2016.40} {\bibfield  {journal} {\bibinfo
  {journal} {Publ. Astron. Soc. Austral.}\ }\textbf {\bibinfo {volume} {33}},\
  \bibinfo {pages} {e048} (\bibinfo {year} {2016})},\ \Eprint
  {https://arxiv.org/abs/1608.03274} {arXiv:1608.03274 [astro-ph.SR]}
  \BibitemShut {NoStop}%
\bibitem [{\citenamefont {Radice}\ \emph {et~al.}(2018)\citenamefont {Radice},
  \citenamefont {Abdikamalov}, \citenamefont {Ott}, \citenamefont {Mosta},
  \citenamefont {Couch},\ and\ \citenamefont {Roberts}}]{Radice:2017kmj}%
  \BibitemOpen
  \bibfield  {author} {\bibinfo {author} {\bibfnamefont {D.}~\bibnamefont
  {Radice}}, \bibinfo {author} {\bibfnamefont {E.}~\bibnamefont {Abdikamalov}},
  \bibinfo {author} {\bibfnamefont {C.~D.}\ \bibnamefont {Ott}}, \bibinfo
  {author} {\bibfnamefont {P.}~\bibnamefont {Mosta}}, \bibinfo {author}
  {\bibfnamefont {S.~M.}\ \bibnamefont {Couch}},\ and\ \bibinfo {author}
  {\bibfnamefont {L.~F.}\ \bibnamefont {Roberts}},\ }\bibfield  {title}
  {\bibinfo {title} {{Turbulence in Core-Collapse Supernovae}},\ }\href
  {https://doi.org/10.1088/1361-6471/aab872} {\bibfield  {journal} {\bibinfo
  {journal} {J. Phys.}\ }\textbf {\bibinfo {volume} {G45}},\ \bibinfo {pages}
  {053003} (\bibinfo {year} {2018})},\ \Eprint
  {https://arxiv.org/abs/1710.01282} {arXiv:1710.01282 [astro-ph.HE]}
  \BibitemShut {NoStop}%
\bibitem [{\citenamefont {Yamamoto}(2016)}]{Yamamoto:2015gzz}%
  \BibitemOpen
  \bibfield  {author} {\bibinfo {author} {\bibfnamefont {N.}~\bibnamefont
  {Yamamoto}},\ }\bibfield  {title} {\bibinfo {title} {{Chiral transport of
  neutrinos in supernovae: Neutrino-induced fluid helicity and helical plasma
  instability}},\ }\href {https://doi.org/10.1103/PhysRevD.93.065017}
  {\bibfield  {journal} {\bibinfo  {journal} {Phys. Rev. D}\ }\textbf {\bibinfo
  {volume} {93}},\ \bibinfo {pages} {065017} (\bibinfo {year} {2016})},\
  \Eprint {https://arxiv.org/abs/1511.00933} {arXiv:1511.00933 [astro-ph.HE]}
  \BibitemShut {NoStop}%
\bibitem [{\citenamefont {Yamamoto}\ and\ \citenamefont
  {Yang}(2020)}]{Yamamoto:2020zrs}%
  \BibitemOpen
  \bibfield  {author} {\bibinfo {author} {\bibfnamefont {N.}~\bibnamefont
  {Yamamoto}}\ and\ \bibinfo {author} {\bibfnamefont {D.-L.}\ \bibnamefont
  {Yang}},\ }\bibfield  {title} {\bibinfo {title} {{Chiral Radiation Transport
  Theory of Neutrinos}},\ }\href {https://doi.org/10.3847/1538-4357/ab8468}
  {\bibfield  {journal} {\bibinfo  {journal} {Astrophys. J.}\ }\textbf
  {\bibinfo {volume} {895}},\ \bibinfo {pages} {56} (\bibinfo {year} {2020})},\
  \Eprint {https://arxiv.org/abs/2002.11348} {arXiv:2002.11348 [astro-ph.HE]}
  \BibitemShut {NoStop}%
\bibitem [{\citenamefont {Yamamoto}\ and\ \citenamefont
  {Yang}(2021{\natexlab{a}})}]{Yamamoto:2021hjs}%
  \BibitemOpen
  \bibfield  {author} {\bibinfo {author} {\bibfnamefont {N.}~\bibnamefont
  {Yamamoto}}\ and\ \bibinfo {author} {\bibfnamefont {D.-L.}\ \bibnamefont
  {Yang}},\ }\bibfield  {title} {\bibinfo {title} {{Magnetic field induced
  neutrino chiral transport near equilibrium}},\ }\href
  {https://doi.org/10.1103/PhysRevD.104.123019} {\bibfield  {journal} {\bibinfo
   {journal} {Phys. Rev. D}\ }\textbf {\bibinfo {volume} {104}},\ \bibinfo
  {pages} {123019} (\bibinfo {year} {2021}{\natexlab{a}})},\ \Eprint
  {https://arxiv.org/abs/2103.13159} {arXiv:2103.13159 [hep-ph]} \BibitemShut
  {NoStop}%
\bibitem [{\citenamefont {Joyce}\ and\ \citenamefont
  {Shaposhnikov}(1997)}]{Joyce:1997uy}%
  \BibitemOpen
  \bibfield  {author} {\bibinfo {author} {\bibfnamefont {M.}~\bibnamefont
  {Joyce}}\ and\ \bibinfo {author} {\bibfnamefont {M.~E.}\ \bibnamefont
  {Shaposhnikov}},\ }\bibfield  {title} {\bibinfo {title} {{Primordial magnetic
  fields, right-handed electrons, and the Abelian anomaly}},\ }\href
  {https://doi.org/10.1103/PhysRevLett.79.1193} {\bibfield  {journal} {\bibinfo
   {journal} {Phys. Rev. Lett.}\ }\textbf {\bibinfo {volume} {79}},\ \bibinfo
  {pages} {1193} (\bibinfo {year} {1997})},\ \Eprint
  {https://arxiv.org/abs/astro-ph/9703005} {arXiv:astro-ph/9703005}
  \BibitemShut {NoStop}%
\bibitem [{\citenamefont {Akamatsu}\ and\ \citenamefont
  {Yamamoto}(2013)}]{Akamatsu:2013pjd}%
  \BibitemOpen
  \bibfield  {author} {\bibinfo {author} {\bibfnamefont {Y.}~\bibnamefont
  {Akamatsu}}\ and\ \bibinfo {author} {\bibfnamefont {N.}~\bibnamefont
  {Yamamoto}},\ }\bibfield  {title} {\bibinfo {title} {{Chiral Plasma
  Instabilities}},\ }\href {https://doi.org/10.1103/PhysRevLett.111.052002}
  {\bibfield  {journal} {\bibinfo  {journal} {Phys. Rev. Lett.}\ }\textbf
  {\bibinfo {volume} {111}},\ \bibinfo {pages} {052002} (\bibinfo {year}
  {2013})},\ \Eprint {https://arxiv.org/abs/1302.2125} {arXiv:1302.2125
  [nucl-th]} \BibitemShut {NoStop}%
\bibitem [{\citenamefont {Vilenkin}(1980)}]{Vilenkin:1980fu}%
  \BibitemOpen
  \bibfield  {author} {\bibinfo {author} {\bibfnamefont {A.}~\bibnamefont
  {Vilenkin}},\ }\bibfield  {title} {\bibinfo {title} {{Equilibrium parity
  violating current in a magnetic field}},\ }\href
  {https://doi.org/10.1103/PhysRevD.22.3080} {\bibfield  {journal} {\bibinfo
  {journal} {Phys. Rev.}\ }\textbf {\bibinfo {volume} {D22}},\ \bibinfo {pages}
  {3080} (\bibinfo {year} {1980})}\BibitemShut {NoStop}%
\bibitem [{\citenamefont {Nielsen}\ and\ \citenamefont
  {Ninomiya}(1983)}]{Nielsen:1983rb}%
  \BibitemOpen
  \bibfield  {author} {\bibinfo {author} {\bibfnamefont {H.~B.}\ \bibnamefont
  {Nielsen}}\ and\ \bibinfo {author} {\bibfnamefont {M.}~\bibnamefont
  {Ninomiya}},\ }\bibfield  {title} {\bibinfo {title} {{The Adler-Bell-Jackiw
  anomaly and Weyl fermions in a crystal}},\ }\href
  {https://doi.org/10.1016/0370-2693(83)91529-0} {\bibfield  {journal}
  {\bibinfo  {journal} {Phys. Lett.}\ }\textbf {\bibinfo {volume} {B130}},\
  \bibinfo {pages} {389} (\bibinfo {year} {1983})}\BibitemShut {NoStop}%
\bibitem [{\citenamefont {Fukushima}\ \emph {et~al.}(2008)\citenamefont
  {Fukushima}, \citenamefont {Kharzeev},\ and\ \citenamefont
  {Warringa}}]{Fukushima:2008xe}%
  \BibitemOpen
  \bibfield  {author} {\bibinfo {author} {\bibfnamefont {K.}~\bibnamefont
  {Fukushima}}, \bibinfo {author} {\bibfnamefont {D.~E.}\ \bibnamefont
  {Kharzeev}},\ and\ \bibinfo {author} {\bibfnamefont {H.~J.}\ \bibnamefont
  {Warringa}},\ }\bibfield  {title} {\bibinfo {title} {{The Chiral Magnetic
  Effect}},\ }\href {https://doi.org/10.1103/PhysRevD.78.074033} {\bibfield
  {journal} {\bibinfo  {journal} {Phys. Rev.}\ }\textbf {\bibinfo {volume}
  {D78}},\ \bibinfo {pages} {074033} (\bibinfo {year} {2008})},\ \Eprint
  {https://arxiv.org/abs/0808.3382} {arXiv:0808.3382 [hep-ph]} \BibitemShut
  {NoStop}%
\bibitem [{\citenamefont {Ohnishi}\ and\ \citenamefont
  {Yamamoto}(2014)}]{Ohnishi:2014uea}%
  \BibitemOpen
  \bibfield  {author} {\bibinfo {author} {\bibfnamefont {A.}~\bibnamefont
  {Ohnishi}}\ and\ \bibinfo {author} {\bibfnamefont {N.}~\bibnamefont
  {Yamamoto}},\ }\bibfield  {title} {\bibinfo {title} {{Magnetars and the
  Chiral Plasma Instabilities}},\ }\href@noop {} {\  (\bibinfo {year}
  {2014})},\ \Eprint {https://arxiv.org/abs/1402.4760} {arXiv:1402.4760
  [astro-ph.HE]} \BibitemShut {NoStop}%
\bibitem [{\citenamefont {Grabowska}\ \emph {et~al.}(2015)\citenamefont
  {Grabowska}, \citenamefont {Kaplan},\ and\ \citenamefont
  {Reddy}}]{Grabowska:2014efa}%
  \BibitemOpen
  \bibfield  {author} {\bibinfo {author} {\bibfnamefont {D.}~\bibnamefont
  {Grabowska}}, \bibinfo {author} {\bibfnamefont {D.~B.}\ \bibnamefont
  {Kaplan}},\ and\ \bibinfo {author} {\bibfnamefont {S.}~\bibnamefont
  {Reddy}},\ }\bibfield  {title} {\bibinfo {title} {{Role of the electron mass
  in damping chiral plasma instability in Supernovae and neutron stars}},\
  }\href {https://doi.org/10.1103/PhysRevD.91.085035} {\bibfield  {journal}
  {\bibinfo  {journal} {Phys. Rev. D}\ }\textbf {\bibinfo {volume} {91}},\
  \bibinfo {pages} {085035} (\bibinfo {year} {2015})},\ \Eprint
  {https://arxiv.org/abs/1409.3602} {arXiv:1409.3602 [hep-ph]} \BibitemShut
  {NoStop}%
\bibitem [{\citenamefont {Dvornikov}\ and\ \citenamefont
  {Semikoz}(2015)}]{Dvornikov:2014uza}%
  \BibitemOpen
  \bibfield  {author} {\bibinfo {author} {\bibfnamefont {M.}~\bibnamefont
  {Dvornikov}}\ and\ \bibinfo {author} {\bibfnamefont {V.~B.}\ \bibnamefont
  {Semikoz}},\ }\bibfield  {title} {\bibinfo {title} {{Magnetic field
  instability in a neutron star driven by the electroweak electron-nucleon
  interaction versus the chiral magnetic effect}},\ }\href
  {https://doi.org/10.1103/PhysRevD.91.061301} {\bibfield  {journal} {\bibinfo
  {journal} {Phys. Rev. D}\ }\textbf {\bibinfo {volume} {91}},\ \bibinfo
  {pages} {061301} (\bibinfo {year} {2015})},\ \Eprint
  {https://arxiv.org/abs/1410.6676} {arXiv:1410.6676 [astro-ph.HE]}
  \BibitemShut {NoStop}%
\bibitem [{\citenamefont {Sigl}\ and\ \citenamefont
  {Leite}(2016)}]{Sigl:2015xva}%
  \BibitemOpen
  \bibfield  {author} {\bibinfo {author} {\bibfnamefont {G.}~\bibnamefont
  {Sigl}}\ and\ \bibinfo {author} {\bibfnamefont {N.}~\bibnamefont {Leite}},\
  }\bibfield  {title} {\bibinfo {title} {{Chiral Magnetic Effect in
  Protoneutron Stars and Magnetic Field Spectral Evolution}},\ }\href
  {https://doi.org/10.1088/1475-7516/2016/01/025} {\bibfield  {journal}
  {\bibinfo  {journal} {JCAP}\ }\textbf {\bibinfo {volume} {01}},\ \bibinfo
  {pages} {025}},\ \Eprint {https://arxiv.org/abs/1507.04983} {arXiv:1507.04983
  [astro-ph.HE]} \BibitemShut {NoStop}%
\bibitem [{\citenamefont {Yamamoto}\ and\ \citenamefont
  {Yang}(2021{\natexlab{b}})}]{Yamamoto:2021gts}%
  \BibitemOpen
  \bibfield  {author} {\bibinfo {author} {\bibfnamefont {N.}~\bibnamefont
  {Yamamoto}}\ and\ \bibinfo {author} {\bibfnamefont {D.-L.}\ \bibnamefont
  {Yang}},\ }\bibfield  {title} {\bibinfo {title} {{Helical magnetic effect and
  the chiral anomaly}},\ }\href {https://doi.org/10.1103/PhysRevD.103.125003}
  {\bibfield  {journal} {\bibinfo  {journal} {Phys. Rev. D}\ }\textbf {\bibinfo
  {volume} {103}},\ \bibinfo {pages} {125003} (\bibinfo {year}
  {2021}{\natexlab{b}})},\ \Eprint {https://arxiv.org/abs/2103.13208}
  {arXiv:2103.13208 [hep-th]} \BibitemShut {NoStop}%
\bibitem [{\citenamefont {Masada}\ \emph {et~al.}(2018)\citenamefont {Masada},
  \citenamefont {Kotake}, \citenamefont {Takiwaki},\ and\ \citenamefont
  {Yamamoto}}]{Masada:2018swb}%
  \BibitemOpen
  \bibfield  {author} {\bibinfo {author} {\bibfnamefont {Y.}~\bibnamefont
  {Masada}}, \bibinfo {author} {\bibfnamefont {K.}~\bibnamefont {Kotake}},
  \bibinfo {author} {\bibfnamefont {T.}~\bibnamefont {Takiwaki}},\ and\
  \bibinfo {author} {\bibfnamefont {N.}~\bibnamefont {Yamamoto}},\ }\bibfield
  {title} {\bibinfo {title} {{Chiral magnetohydrodynamic turbulence in
  core-collapse supernovae}},\ }\href
  {https://doi.org/10.1103/PhysRevD.98.083018} {\bibfield  {journal} {\bibinfo
  {journal} {Phys. Rev. D}\ }\textbf {\bibinfo {volume} {98}},\ \bibinfo
  {pages} {083018} (\bibinfo {year} {2018})},\ \Eprint
  {https://arxiv.org/abs/1805.10419} {arXiv:1805.10419 [astro-ph.HE]}
  \BibitemShut {NoStop}%
\bibitem [{\citenamefont {Brandenburg}\ \emph {et~al.}(2017)\citenamefont
  {Brandenburg}, \citenamefont {Schober}, \citenamefont {Rogachevskii},
  \citenamefont {Kahniashvili}, \citenamefont {Boyarsky}, \citenamefont
  {Frohlich}, \citenamefont {Ruchayskiy},\ and\ \citenamefont
  {Kleeorin}}]{Brandenburg:2017rcb}%
  \BibitemOpen
  \bibfield  {author} {\bibinfo {author} {\bibfnamefont {A.}~\bibnamefont
  {Brandenburg}}, \bibinfo {author} {\bibfnamefont {J.}~\bibnamefont
  {Schober}}, \bibinfo {author} {\bibfnamefont {I.}~\bibnamefont
  {Rogachevskii}}, \bibinfo {author} {\bibfnamefont {T.}~\bibnamefont
  {Kahniashvili}}, \bibinfo {author} {\bibfnamefont {A.}~\bibnamefont
  {Boyarsky}}, \bibinfo {author} {\bibfnamefont {J.}~\bibnamefont {Frohlich}},
  \bibinfo {author} {\bibfnamefont {O.}~\bibnamefont {Ruchayskiy}},\ and\
  \bibinfo {author} {\bibfnamefont {N.}~\bibnamefont {Kleeorin}},\ }\bibfield
  {title} {\bibinfo {title} {{The turbulent chiral-magnetic cascade in the
  early universe}},\ }\href {https://doi.org/10.3847/2041-8213/aa855d}
  {\bibfield  {journal} {\bibinfo  {journal} {Astrophys. J. Lett.}\ }\textbf
  {\bibinfo {volume} {845}},\ \bibinfo {pages} {L21} (\bibinfo {year}
  {2017})},\ \Eprint {https://arxiv.org/abs/1707.03385} {arXiv:1707.03385
  [astro-ph.CO]} \BibitemShut {NoStop}%
\bibitem [{\citenamefont {Schober}\ \emph {et~al.}(2018)\citenamefont
  {Schober}, \citenamefont {Rogachevskii}, \citenamefont {Brandenburg},
  \citenamefont {Boyarsky}, \citenamefont {Fr\"ohlich}, \citenamefont
  {Ruchayskiy},\ and\ \citenamefont {Kleeorin}}]{Schober:2017cdw}%
  \BibitemOpen
  \bibfield  {author} {\bibinfo {author} {\bibfnamefont {J.}~\bibnamefont
  {Schober}}, \bibinfo {author} {\bibfnamefont {I.}~\bibnamefont
  {Rogachevskii}}, \bibinfo {author} {\bibfnamefont {A.}~\bibnamefont
  {Brandenburg}}, \bibinfo {author} {\bibfnamefont {A.}~\bibnamefont
  {Boyarsky}}, \bibinfo {author} {\bibfnamefont {J.}~\bibnamefont
  {Fr\"ohlich}}, \bibinfo {author} {\bibfnamefont {O.}~\bibnamefont
  {Ruchayskiy}},\ and\ \bibinfo {author} {\bibfnamefont {N.}~\bibnamefont
  {Kleeorin}},\ }\bibfield  {title} {\bibinfo {title} {{Laminar and turbulent
  dynamos in chiral magnetohydrodynamics. II. Simulations}},\ }\href
  {https://doi.org/10.3847/1538-4357/aaba75} {\bibfield  {journal} {\bibinfo
  {journal} {Astrophys. J.}\ }\textbf {\bibinfo {volume} {858}},\ \bibinfo
  {pages} {124} (\bibinfo {year} {2018})},\ \Eprint
  {https://arxiv.org/abs/1711.09733} {arXiv:1711.09733 [physics.flu-dyn]}
  \BibitemShut {NoStop}%
\bibitem [{\citenamefont {{Miyoshi}}\ and\ \citenamefont
  {{Kusano}}(2005)}]{Miyoshi05}%
  \BibitemOpen
  \bibfield  {author} {\bibinfo {author} {\bibfnamefont {T.}~\bibnamefont
  {{Miyoshi}}}\ and\ \bibinfo {author} {\bibfnamefont {K.}~\bibnamefont
  {{Kusano}}},\ }\bibfield  {title} {\bibinfo {title} {{A multi-state HLL
  approximate Riemann solver for ideal magnetohydrodynamics}},\ }\href
  {https://doi.org/10.1016/j.jcp.2005.02.017} {\bibfield  {journal} {\bibinfo
  {journal} {Journal of Computational Physics}\ }\textbf {\bibinfo {volume}
  {208}},\ \bibinfo {pages} {315} (\bibinfo {year} {2005})}\BibitemShut
  {NoStop}%
\bibitem [{\citenamefont {{Evans}}\ and\ \citenamefont
  {{Hawley}}(1988)}]{Evans88}%
  \BibitemOpen
  \bibfield  {author} {\bibinfo {author} {\bibfnamefont {C.~R.}\ \bibnamefont
  {{Evans}}}\ and\ \bibinfo {author} {\bibfnamefont {J.~F.}\ \bibnamefont
  {{Hawley}}},\ }\bibfield  {title} {\bibinfo {title} {{Simulation of
  Magnetohydrodynamic Flows: A Constrained Transport Model}},\ }\href
  {https://doi.org/10.1086/166684} {\bibfield  {journal} {\bibinfo  {journal}
  {\apj}\ }\textbf {\bibinfo {volume} {332}},\ \bibinfo {pages} {659} (\bibinfo
  {year} {1988})}\BibitemShut {NoStop}%
\end{thebibliography}%
\end{document}